\documentclass[twocolumn]{article}

\usepackage[margin=1in]{geometry} 

\usepackage{microtype}
\usepackage{natbib}
\usepackage{graphicx}
\usepackage{stackengine}
\usepackage{subcaption}
\usepackage[table]{xcolor}
\usepackage{booktabs}

\usepackage{hyperref}

\usepackage{amsmath,amssymb,amsthm,mathtools}
\usepackage{dcolumn}
\usepackage{quantikz}
\usepackage{makecell}
\usepackage{physics}
\usepackage{multirow}
\usepackage{orcidlink}
\usepackage{bbm}
\usepackage{tikz}
\usepackage[disable]{todonotes}

\renewcommand{\arraystretch}{2}

\theoremstyle{plain}

\theoremstyle{definition}

\theoremstyle{remark}

\title{Reinforcement learning with learned gadgets to tackle hard quantum problems on real hardware}

\author{
  Akash Kundu\orcidlink{0000-0002-3540-1061}\thanks{Department of Physics, University of Helsinki, Helsinki, Finland. Email: \texttt{akash.kundu@helsinki.fi}}
  \and
  Leopoldo Sarra\orcidlink{0000-0001-7504-8656}\thanks{Axiomatic AI, Barcelona, Spain}
}

\date{} 

\begin{document}

\maketitle

\begin{abstract}
    Quantum computing offers exciting opportunities for simulating complex quantum systems and optimizing large-scale combinatorial problems, but its practical use is limited by device noise and constrained connectivity. Designing quantum circuits, which are fundamental to quantum algorithms, is therefore a central challenge in current quantum hardware. Existing reinforcement learning‑based methods for circuit design lose accuracy when restricted to hardware‑native gates and device‑level compilation. Here, we introduce gadget reinforcement learning (GRL) that combines learning with program synthesis to automatically construct composite gates that expand the action space while respecting hardware constraints. We show that this approach improves accuracy, hardware compatibility, and scalability for transverse field Ising and quantum chemistry problems, reaching systems of up to ten qubits within realistic computational budgets. This framework demonstrates how learned, reusable circuit building blocks can guide the co‑design of algorithms and hardware for quantum processors.
\end{abstract}


\section{Introduction}
\label{sec:introduction}

Quantum computing has experienced substantial advancements in recent years, unlocking the potential to solve classically intractable problems. Foundational algorithms like Shor's algorithm for integer factorization~\cite{shor1999polynomial} and Grover's algorithm for unstructured search~\cite{grover1996fast} demonstrate the transformative promise of quantum technology. However, practical implementation of these algorithms faces substantial hurdles due to the limitations of current quantum hardware, characterized by small qubit counts, significant noise, and constrained connectivity~\cite{monz2016realization, mandviwalla2018implementing}. These challenges require innovative approaches to bridge the gap between theoretical breakthroughs and hardware capabilities.
\begin{figure*}[t]
    \centering
    \includegraphics[width=0.9\linewidth]{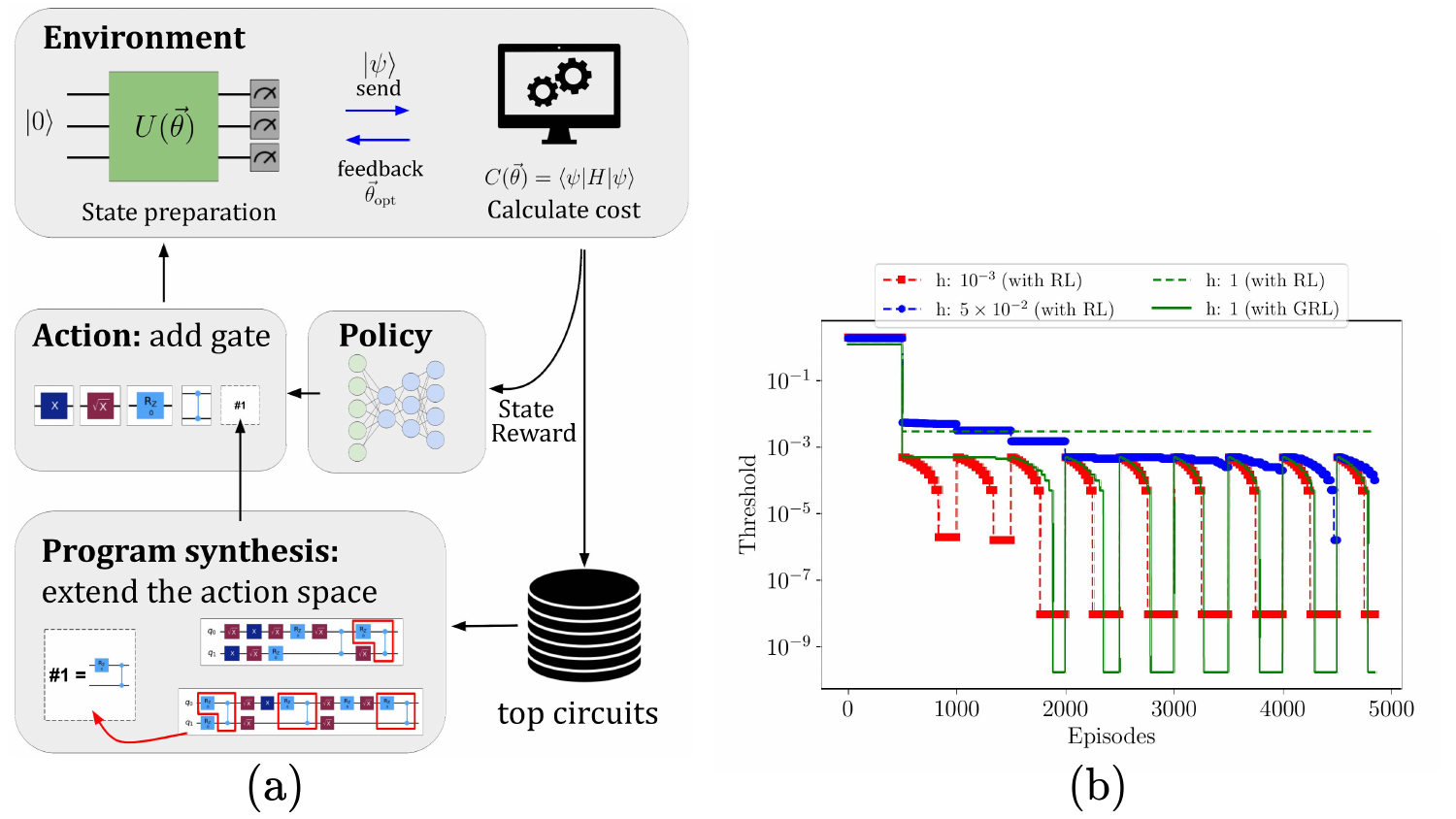}
    \caption{
    {Gadget reinforcement learning (GRL) framework for quantum circuit construction and ground state preparation. (a) GRL algorithm:}
    An RL agent sequentially builds a quantum circuit for state preparation, using the energy expectation of a Hamiltonian as the cost.
    Rewards $\pm r$ are assigned based on whether the cost falls below a threshold $\zeta$, guiding policy updates.
    The top-$k$ circuits are analyzed via program synthesis to extract composite gates ({gadgets}), which are added to the action space for further training.
    {(b) GRL vs. RL~\cite{patel2024curriculum} on the transverse field Ising model (TFIM):}
    Considering a 2-qubit TFIM (Eq.~\ref{eq:tfim_model}) with varying field strength $h$, standard RL fails to find the ground state for $h=1$, while GRL maintains high accuracy as $h$ increases.}
    \label{fig:GRL_algorithm_sketch}
\end{figure*}

Hybrid quantum-classical algorithms, particularly variational quantum algorithms (VQAs), have emerged as a promising route to exploit near-term quantum devices. VQAs operate by dividing computation between quantum hardware and classical optimization. Their implementation involves three main steps: (1) {Quantum state preparation}: A parameterized quantum circuit (PQC) \(U(\vec{\theta})\), containing adjustable parameters \(\vec{\theta}\), is constructed using single-qubit rotations and typically non-parameterized two-qubit entangling gates. (2) {Measurement}: The PQC is executed on quantum hardware to evaluate the cost function
\begin{equation}
   C(\vec{\theta}) = \langle 0 | U^\dagger(\vec{\theta}) H U(\vec{\theta}) | 0 \rangle,
   \label{eq:cost_function_vqa}
\end{equation}
where \(H\) represents the Hamiltonian encoding the problem. (3) {Optimization}: Classical algorithms minimize \(C(\vec{\theta})\) by adjusting \(\vec{\theta}\). This paradigm transforms solving a quantum problem into designing hardware-efficient PQCs that achieve low cost.

However, designing effective PQCs remains difficult due to the constraints of current quantum hardware. Different noise levels, qubit connectivity topologies, and gate fidelities complicate the process, making hardware-specific PQC design particularly challenging. Recent efforts have focused on adaptive ansatz construction~\cite{grimsley2019adaptive, tang2021qubit, feniou2023overlap} and advanced optimization techniques~\cite{zhou2020quantum, zhu2022adaptive, cheng2024quantum, kundu2024hamiltonian} to mitigate barren plateaus and improve trainability. In parallel, machine learning approaches, in particular reinforcement learning (RL), have emerged as powerful tools for automating PQC design and quantum control~\cite{FM_REVIEW_PhysRevA.107.010101, bang2014strategy, ostaszewski2021reinforcement, kundu2024reinforcementqas}.

RL-based methods have been applied to ground-state preparation, entanglement generation, and quantum machine learning, using deep-Q network~\cite{mnih2013playing} agents with tailored reward functions~\cite{ostaszewski2021reinforcement, ye2021quantum}. Other works leverage deep RL for hardware-aware circuit optimization and routing~\cite{fosel2021quantum, patel2024curriculum, tang2024alpharouter}, as well as for automatically generating ansatz families utilizing cost explosion for quantum algorithms~\cite{moflic2023cost}. Recent benchmarks highlight that RL for quantum circuit design faces persistent challenges, including sparse rewards, hybrid discrete--continuous action spaces, and large fixed action sets~\cite{patel2024reinforcement, sadhu2024quantumqas, altmann2024challenges}.

Program synthesis offers a complementary angle by extracting high-level structure from low-level primitives. Classical syntax-guided synthesis (SyGuS)~\cite{alur2013syntax} and proof-theoretic approaches have inspired quantum program synthesis frameworks for unitary decomposition and parameterized circuit construction~\cite{srivastava2010program, deng2024case, DiscoveringQuantumCircuits_Sarra_2024}. Inspired by systems such as DreamCoder~\cite{DREAMCODER_Ellis2020DreamCoderGG} and compression-based enumeration~\cite{ENUMERATION_COMPRESSION_10.5555/2540128.2540316}, these methods show that analyzing simpler tasks to extract reusable fragments can reduce search complexity. In the quantum domain, composite gates or ``gadgets'' have been proposed to compress circuits and aid interpretability~\cite{GADGETS_Ruiz:2024boz, GADGETS_Trenkwalder_2023}, but their iterative use to enhance RL agents remains largely unexplored.

In conventional RL-based approaches, an agent explores a fixed action space comprising predefined quantum gates to construct PQCs. While effective for a set of quantum optimization problems, a fixed gate set limits adaptability and scalability: as the number of qubits or the problem hardness grows, agents require extensive exploration and often suffer performance degradation under realistic computational budgets. Curriculum RL partially alleviates this by presenting tasks of increasing difficulty~\cite{patel2024curriculum}, yet still operates with a static, low-level action space that does not capture emerging higher-level structure. Moreover, these PQCs further require transpilation to a quantum hardware native gateset, which degrades the performance of the trained PQC.

We propose to address parts of this challenge with a framework capable of leveraging insights from simpler problems to solve more complex ones efficiently, within a fixed computational budget. This paper introduces {gadget reinforcement learning} (GRL), an approach that combines RL with program synthesis to dynamically expand the agent’s action space. GRL achieves this by synthesizing higher-level composite gates, or ``gadgets'', from the solutions of simpler problem instances and incorporating these gadgets into the agent's action space. By doing so, GRL enhances the agent’s ability to generalize and adapt, optimizing computational resource utilization.

A schematic of the GRL algorithm is presented in Figure~\ref{fig:GRL_algorithm_sketch}(a). The process begins with an RL agent solving a simple instance of a problem using a basic action space, such as the native gateset of a specific quantum processor. The program synthesis component identifies recurring patterns in the top-performing circuits, synthesizes these patterns into gadgets, and adds them to the action space. With this expanded action space, the RL agent retrains to tackle more challenging problem instances.

To demonstrate the efficacy of GRL, we apply it to the transverse field Ising model (TFIM), a problem that becomes increasingly difficult as the magnetic field strength \(h\) or the system size grows. GRL learns gadgets from a simple 2-qubit TFIM with \(h = 10^{-3}\) and successfully uses them to solve more complex instances, including a 3-qubit TFIM at \(h = 1\), a regime where conventional RL approaches fail due to computational limitations. The comparison of performance across regimes is shown in Figure~\ref{fig:GRL_algorithm_sketch}(b).

Our results highlight the advantages of GRL: (1) {Improved computational efficiency}: By learning and leveraging gadgets, GRL achieves superior performance within a fixed computational budget, avoiding the exhaustive exploration required in fixed-action-space RL. (2) {Scalability through gadget transfer}: GRL effectively generalizes knowledge from simpler tasks to more complex ones, reducing the computational burden associated with solving larger problems up to 10-qubit. (3) {Hardware compatibility}: The PQCs generated by GRL are compact and directly employ the \texttt{IBMQ Heron} processor native gateset, i.e., \texttt{\{RZ, SX, X, CZ\}}. In addition, they are connectivity-optimized, making them more resilient under noise and practical for real-world implementation. (4) {GRL is effective beyond TFIM}: Here, we consider the learned gadgets during the ground state preparation of 2-qubit H$_2$ and utilize these gadgets in training 3-qubit H$_2$ molecule with GRL. The analysis shows that extending the operators in the action space with gadgets helps us achieve lower error than training without them.

\section{Methods}
\label{sec:methods}

Here, the goal is to construct parameterized quantum circuits (PQCs) that efficiently solve families of quantum optimization problems. The central idea is to let a reinforcement learning (RL) agent explore the PQC design space while a program synthesis routine distills frequently used gate patterns into reusable composite operations, which we refer to as ``gadgets". These gadgets are then added back into the action space, enabling the agent to reason in terms of higher-level building blocks rather than only elementary gates, and thereby accelerate the search for effective circuits on more challenging instances.

In the remainder of this section, the focus is on three components of this framework. First, the gadget reinforcement learning (GRL) algorithm is introduced, describing how the RL agent constructs PQCs, how circuits are encoded as tensors, and how rewards guide the agent toward low-energy solutions in a VQA setting. Second, the library-building module is presented, which uses a program synthesis approach to analyze high-performing circuits and extract a library of gadgets that augment the hardware-native gate set. Third, these ingredients are instantiated on the transverse field Ising model (TFIM), illustrating how gadgets learned on simpler parameter regimes can be reused to solve progressively harder instances with improved sample efficiency and circuit compactness.

\subsection{Gadget reinforcement learning}\label{sec:grl_algorithm}
We provide an overview of the GRL algorithm for constructing PQCs in a VQA task. Consequently, we provide details on the state and action representations as well as the reward function employed in this study.

The GRL algorithm initiates with an empty quantum circuit. The RL agent, based on a double deep Q-network and $\epsilon$-greedy policy (for further details, see {Supplementary Note 4.2 in Supplementary Information}), sequentially appends the gates to the circuit until the maximum number of actions has been reached.
The actions are chosen from an action space of available elementary gates. In particular, our application contains \texttt{RZ}, \texttt{SX}, \texttt{X} as 1-qubit gates, where \texttt{RZ} is the only parameterized gate in the action space.
Furthermore, to entangle the qubits, we use a controlled-Z (\texttt{CZ}) gate.
The main reason for choosing such an action space is that all these gates are the native gates of the \texttt{IBM Heron} processor. 
Therefore, we do not need to further transpile the circuits, which is itself an NP-hard task~\cite{wille2023mqt, botea2018complexity}, when executing on the processor.
We implement a double-deep RL method, where the PQCs are encoded in a refined binary tensor representation, as proposed in~\cite{kundu2024kanqas}.
This encoding is inspired by the tensor-based encoding proposed in~\cite{patel2024curriculum} as illustrated in Figure~\ref{fig:tensor_based_encoding}. In the Supplementary Note 4.1 in Supplementary Information, we thoroughly describe the refined encoding scheme with an example.
\begin{figure}[h!]
    \centering
    \includegraphics[width=\linewidth]{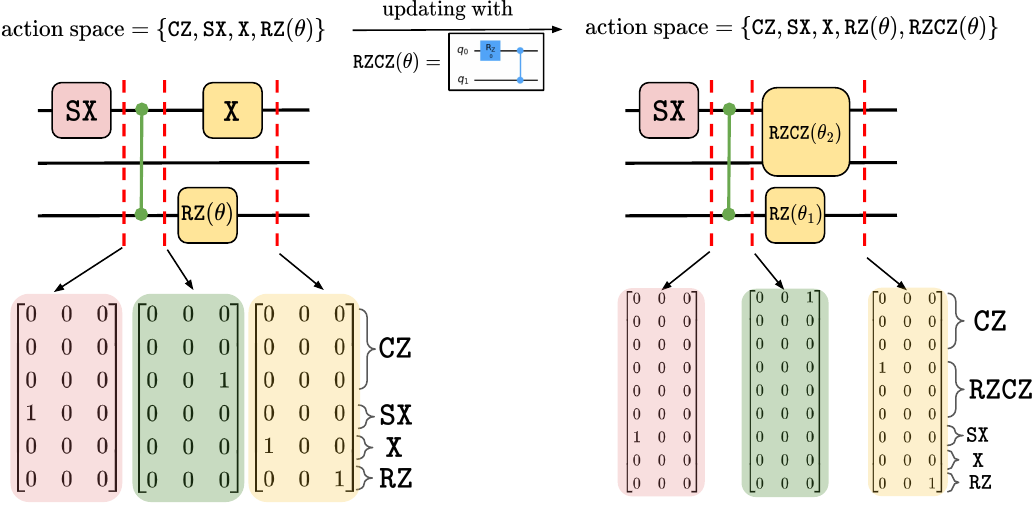}
    \caption{Illustration of the gadget-informed tensor encoding of parameterized quantum circuits for reinforcement learning. This is the RL-state for the reinforcement learning algorithm. The 2-qubit gates are encoded into a matrix whose dimension is dependent on the number of qubits. Meanwhile, the 1-qubit gates are encoded into the remaining $N_\text{1q}$ rows, which define the number of different 1-qubit gates present in the action space. 
    After the program synthesis algorithm finds the most common patterns of gates, i.e. {gadgets}, in the top performing PQCs, the action space is then updated with the extracted {gadget}. 
    In the gadget reinforcement learning, the dimension of the tensor is then increased. The increase in dimension depends on whether the {gadget} is a 1- or 2-qubit gate.}
    \label{fig:tensor_based_encoding}
\end{figure}

To steer the agent towards the target, we use the same reward function $R$ at every time step $t$ of an episode, as in~\cite{ostaszewski2021reinforcement} defined by
\begin{equation}
R = 
\begin{cases} 
r, & \text{if } C_t < \zeta, \\
-r & \text { if } t \geq T_\text{max} \text { and } C_t \geq \zeta, \\
\mathcal{C}, & \text{otherwise}.
\end{cases}
\label{case:error_rwd}
\end{equation}
where $\mathcal{C} = \text{max}\left(\frac{C_{t-1} - C_t}{\lvert C_{t-1} - C_\text{min} \rvert}, -1\right)$, \( r \) is a real positive number, \( C_t \) {represents} the value of the cost function \( C \) (as defined in Eq.~\ref{eq:cost_function_vqa}) at step \( t \), and \( T_\text{max} \) {denotes} the maximum number of steps allowed for an episode. Additionally, note that when the agent receives a positive reward value \( r \), the episode concludes. In other words, there are two stopping conditions: either surpassing the threshold \( \zeta \) or reaching the maximum number of actions. The agent's objective is to estimate the value \( C_\text{min} \) with the desired precision $\zeta$.

In what follows, we utilize a feedback-driven curriculum reinforcement learning agent.
In particular, the agent updates its threshold while running the episodes: if we find a ground state with lower energy than the threshold, we decrease the threshold; otherwise, we increase it again.
The algorithm is described with more technical detail in Supplementary Note 2 in the Supplementary Information. 

\subsection{Library building}
\label{sec:library-algorithm}

In the next step, we sample the top $k$ PQCs,  chosen according to how effective they are at estimating the solution to the problems, i.e., with a smaller associated $\zeta$ value. These PQCs are then processed through a program synthesis algorithm. {This algorithm can extract composite gates, i.e., {gadgets}, by choosing those with the largest log-likelihood. In particular, it strikes a balance between the expected usage frequency and the gate sequence length. This encourages the extraction of short gate sequences that are expected to be used often}. We refer the reader to the Supplementary Note 3 in Supplementary Information for more technical details on how the likelihood is estimated.
We construct the GRL algorithm by updating the action space with the {gadgets} discovered by the library building module. Finally, the GRL is executed again with the modified action space, consisting of the initial gateset corresponding to the quantum hardware and the {gadgets}.

To update the action space in GRL, a library-building algorithm that leverages a program synthesis framework inspired by~\cite{DiscoveringQuantumCircuits_Sarra_2024, DREAMCODER_Ellis2020DreamCoderGG} is employed.
{The algorithm analyzes the top-$k$ PQCs to identify and extract recurrent gate sequences}. The PQCs are expressed as programs in a typed-$\lambda$-calculus formalism~\cite{LAMBDA_pierce_types_2002}, where the gates act as functions that take a quantum circuit and the target qubits as inputs and return the updated PQC with the gate applied.
For example, a function that applies an $X$ gate on the first qubit and then a controlled-$Z$ gate can be represented as
\begin{equation}
    f(I_2) = cz(x(I_2,0),0,1)
\end{equation}
where $I_2$ is a 2-qubit empty circuit.
Each circuit program is organized into a syntax tree.
The algorithm decomposes each circuit into fragments, i.e. sets of operations, and looks for the most common fragments in the input set.
We use the fragment grammar formalism to evaluate each fragment's usefulness based on a grammar score.
In this context, a grammar $g$ consists of elementary gates (primitives) with usage probabilities estimated from the given set of $k$ top circuits. The grammar score function prioritizes grammars that are most likely to effectively produce the given set of circuits while balancing complexity.

We then modify the action space of the RL agent by adding the highest-scoring fragments, which are expected to help find more compact PQCs with a smaller number of gates.
In our experiments, we show that, although the library is built upon problems that are small and simple, these libraries generalize effectively and can be utilized to GRL and solve harder instances of the given problem iteratively.
For further details on grammar scoring, fragment grammar structure, and hyperparameter settings, refer to Supplementary Note 3 in Supplementary Information.

The GRL runs iteratively by first considering a small system (e.g. in our case a 2-qubit Ising model in a weak transverse field, $h=10^{-3}$) and finding the solution within a pre-defined threshold ($\zeta$).
The agent then finds the ground state within the compute budget, expressed by a fixed number of episodes.
Subsequently, we try to solve an intermediate difficulty problem (in our case, the Ising model with a larger transverse field, $h=5\times10^{-2}$).

As an example application for our algorithm, we consider the transverse field Ising model (TFIM).
The goal is to design a circuit that finds the system's ground state, i.e., the system with the lowest energy.
{Estimating the ground state of TFIM is a complete problem for the complexity class StoqMA, which is an extension of the classical class MA}~\cite{bravyi2017complexity}.
The Hamiltonian of the system is defined by
\begin{equation}
    H = -J \sum_{\langle i, j \rangle}^N \sigma_i^z \sigma_j^z - h \sum_{i} \sigma_i^x
    \label{eq:tfim_model}
\end{equation}

where $N$ is the number of qubits, \( J \) is the coupling constant between neighboring spins, \( h \) is the strength of the transverse field, \( \sigma_i^z \) and \( \sigma_i^x \) are the Pauli matrices acting on the \(i\)-th spin in the \(z\)- and \(x\)-direction, respectively, and \( \langle i, j \rangle \) denotes summation over nearest neighbors.
This model shows a ferromagnetic phase transition at $J\gg h$ and has been studied thoroughly in the literature, for example, with hybrid quantum-classical approaches~\cite{TFIM_SOTA_Sumeet:2023awz} where they utilize numerical linked-cluster expansions with the variational quantum eigensolver (VQE) for TFIM with one-dimensional chains and the two-dimensional square lattice.
\begin{figure*}[t!]
    \centering
    \includegraphics[width=\linewidth]{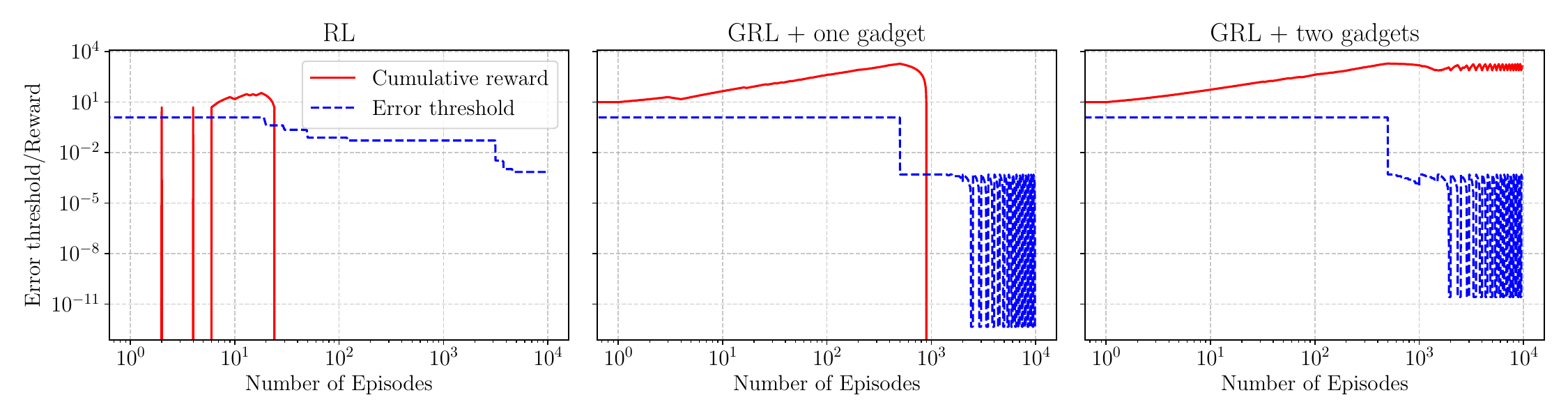}
    \caption{{Cumulative reward improvement of gadget reinforcement learning (GRL) over RL for preparing the ground state of TFIM. We consider $N=2$ TFIM, GRL with 1- and 2-gadgets (see the title of each figure)improves the cumulative reward (red) growth compared to RL}. The RL-only agent struggles to provide improvement in cumulative rewards and returns. GRL with one gadget improves performance, achieving steady cumulative reward growth and frequent positive returns, but experiences a notable drop around the $1000$th episode, reaching the machine precision error threshold (blue). GRL with two gadgets help escalate this drop and reach machine precision with fewer agent-environment interactions.}
    \label{fig:N=2_reward}
\end{figure*}

The primary motivation for using the TFIM in the GRL framework is the increasing difficulty in finding the ground state as the magnetic field strength $h$ varies from small values (on the order of $10^{-3}$) towards $1$, which is defined as the phase change point. This difficulty arises due to the degeneracy between the ground and first excited states that emerge as $h$ approaches the critical value~\cite{curro2024quantum, pfeuty1970one}. In Supplementary Note 5 in Supplementary Information, we show that the degeneracy phenomenon is a key feature of the quantum phase transition in the TFIM and significantly impacts the behavior of the system near the critical point.

The primary objective of employing GRL is to derive {gadgets} from easily solvable instances (in our case, where $h \ll J$) through program synthesis within an RL framework. These gadgets are then incorporated into the action space of the RL agent, enabling more efficient solutions for harder problem instances. We assess this efficiency through two key metrics: (1) agent performance, measured via the cumulative reward, the nature of agent-environment interactions, and the overall training duration; and (2) training accuracy, evaluated by comparing the outcomes of the GRL agent against state-of-the-art RL agents. Our analysis focuses on the number of 1- and 2-qubit gates required to achieve a specified accuracy, both in simulated settings and on actual quantum hardware.  

\section{Results}
In this section we elaborately discuss the performance of gadget reinforcement learning (GRL) in tackling transverse field Ising model (TFIM) and ground state preparation of H$_2$ molecule.
\subsection{Improved performance}

\paragraph{Agent accuracy and success frequency}
Supplementary Figure 4 illustrates the performance of RL and GRL agents in finding the TFIM ground state. The RL-only framework starts with a small system in an easy regime (e.g., weak transverse field, $h=10^{-3}$) and achieves machine precision within a fixed compute budget (up to 48 hours of training). For the intermediate regime ($h=5\times10^{-2}$), the agent finds an approximation, but the PQCs are large, and the errors are relatively high. The RL-agent fails to give us a good approximation of the ground state for $h=1$ and the number of successful episodes (an episode is deemed successful if the agent approximates the ground state within a predefined threshold $\zeta$). drastically reduces as we increase the target precision.  

To improve efficiency, we analyze the top $k$ PQCs from earlier cases and extract key components as new primitive composite gates, or {gadgets}.
While solving the easy instances in the weak magnetic field regime, we identified two key gadgets: $RZ_{i}CZ_{ij}$ (also denoted as $RZCZ$) and $X_{i}\sqrt{X_{i}}$ (also denoted as $X\sqrt{X}$), where $i$ and $j$ denote qubit positions. These gadgets possess desirable properties in terms of both symmetry alignment and experimental feasibility. The $RZCZ$ gadget naturally conforms to the parity symmetry of the TFIM, thereby improving learning efficiency by ensuring the resulting circuits respect conservation laws. Moreover, since both $RZ$ and $CZ$ gates are native to many superconducting qubit architectures, they are highly suited for robust and scalable hardware deployment whose performace does not degrade while transpiled on real quantum hardware.

{In contrast, the $X\sqrt{X}$ gadget does not generally commute with other operations, yet when carefully placed and appropriately repeated, it can collectively preserve the model’s symmetry. This reveals that symmetry preservation is not an inherent property of individual gates but can instead emerge from their structured composition within circuit fragments. Commutativity also plays a role in practical use: although $RZ$ and $CZ$ commute when acting on disjoint qubits, the $X\sqrt{X}$ gadget requires more constrained placement to avoid breaking the global symmetry.}  

{Taken together, these observations highlight that the most practical gadgets are those that strike a balance between aligning with the Hamiltonian’s symmetries, supporting modular circuit synthesis, and being readily implementable on existing quantum hardware. Furthermore, the strong performance of such gadgets may signal the presence of hidden or previously unknown symmetries in the underlying model. In the following, we proceed to show that, as these gadgets respect the Hamiltonian symmetries, we observe significant improvements in the performance of curriculum RL for quantum circuit synthesis in the presence of gadgets.}

By adding the most likely gadgets to the RL agent's action space, we achieve significantly better approximations of the ground state. As additional gadgets are included, the agent experiences increasingly frequent successful episodes, achieving progressively lower errors in estimating the ground state.

We recall that GRL runs iteratively, with the agent and environment specifications; the hyperparameter details are provided in Supplementary Note 4.3 in Supplementary Information. Additionally, in Supplementary Note 10 in Supplementary Information, we give a more detailed analysis of the training time required by both the RL and GRL methods.
Due to the limitations of the available cluster, we restricted the training to 5000 episodes and a maximum of 48 hours of runtime.

\begin{figure*}[h!]
    \centering
    \includegraphics[width=0.7\linewidth]{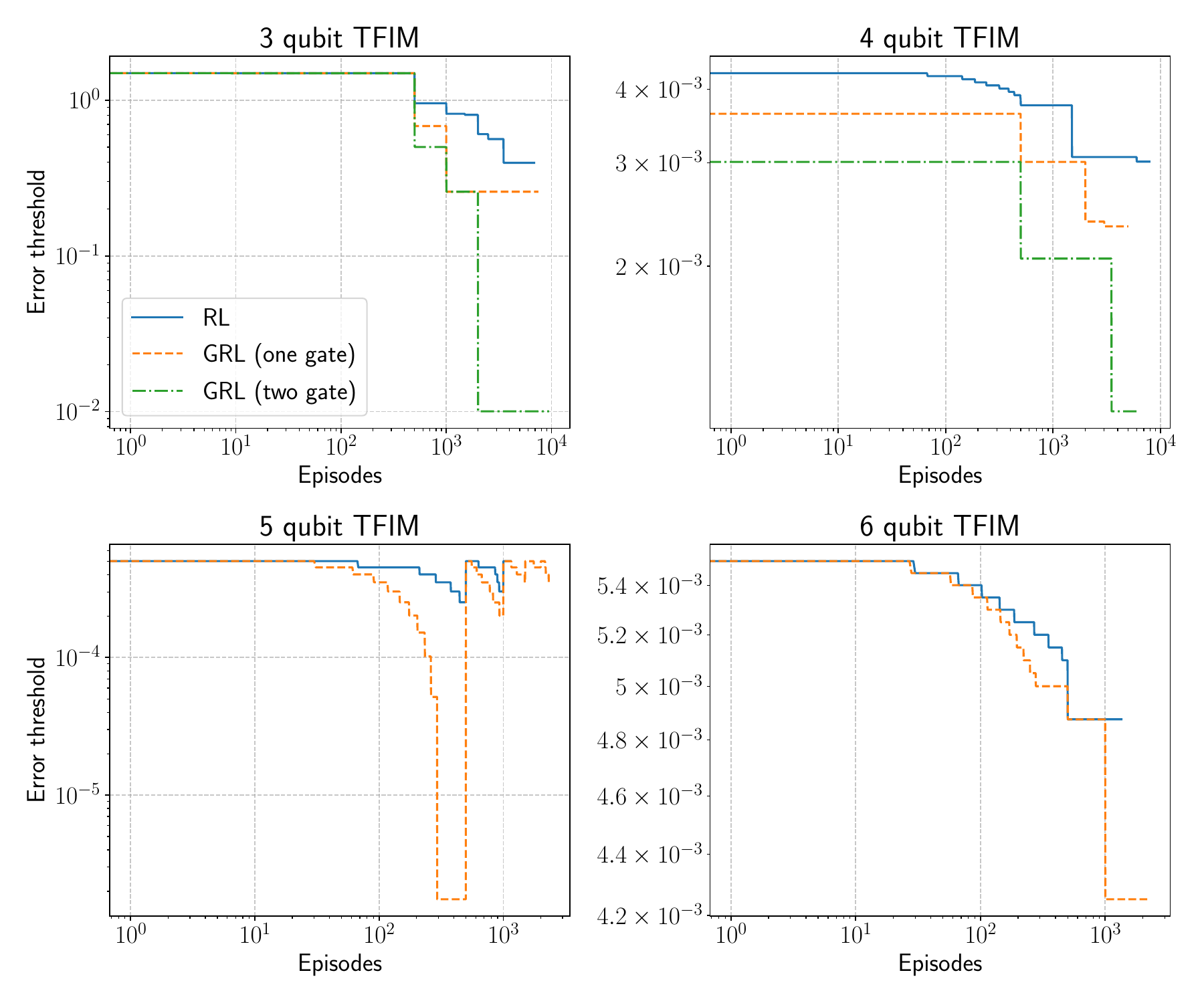}
    \caption{{Gadget reinforcement learning (GRL) performance with increasing system size. Panels show the error threshold versus training episodes for 3-, 4-, 5-, and 6-qubit TFIM instances using standard RL (blue), GRL with 1-gadget (orange), and GRL with 2-gadget (green, where applicable). The gadgets are extracted from 2‑qubit TFIM and then transferred to larger systems. For 3‑qubit TFIM and 4‑qubit TFIM, GRL with gadgets reaches lower error thresholds faster than RL. For 5‑qubit and 6‑qubit TFIM, GRL with a single gadget still outperforms RL, illustrating the scalability of the GRL.}}
    \label{fig:scalability-of-gadgets}
\end{figure*}
\paragraph{Cumulative rewards and returns}
Figure~\ref{fig:N=2_reward} shows that the RL-only agent struggles with consistent rewards and positive returns, while GRL with one gadget steadily improves but plateaus around the $1000$th episode, reaching machine precision. Adding a second gadget overcomes and improves performance. It is worth noting that, for the $N=2$ qubit problem, machine precision can be achieved with just one extracted gadget, making the second gadget redundant in this case. The two-gadget implementation for $N=2$ serves to illustrate the potential for performance improvement when additional gadgets are introduced in more complex systems. A similar observation is recorded for $N=3$ qubit TFIM and is described in Supplementary Note 6 in Supplementary Information.

{
\subsection{GRL against RL and non-RL baselines.}
Compared to a state-of-the-art curriculum-based RL approach~\cite{patel2024curriculum}, the GRL agent is more effective, particularly in harder regimes. Gadget extraction is performed on easier tasks, and the resulting modified action space is used to tackle more challenging problems, such as $h=1$.}

To broaden our evaluation, we also benchmark GRL against established non-RL methods for quantum circuit design for action space \texttt{RZ}, \texttt{X}, \texttt{SX}, and \texttt{CZ}. The baselines include evolutionary algorithms such as QNEAT, the hybrid evolutionary algorithm (Hybrid EA)~\cite{sunkel2025quantum}, and the hardware-efficient ansatzes (HEA)~\cite{kandala2017hardware}. For QNEAT parameters, we follow standard values from previous work: the number of individuals $N = 150$, mutation scale $\sigma = 0.01$, weight mutation probability $p_w = 0.3$, 1-qubit gate mutation probability $p_{\mathrm{1Q}} = 0.5$, \texttt{CZ} mutation probability $p_{\mathrm{CZ}} = 0.5$, $\delta_0 = 1.0$, and $i_L = 0/1/2$. Whereas for the Hybrid EA, we adopt the hyperparameters provided in prior work: population size 200, 1000 generations, crossover and mutation rates each set to 0.85, and offspring and replace rates set to 0.3. For HEA, each layer contains \texttt{RY} gate followed by \texttt{CX} gates connected in a cyclic manner. {In addition, we compare the GRL approach with the constant-depth TFIM square and triangle circuit constructions as proposed in~\cite{camps2022algebraic}, using both circuit layouts as deterministic baselines whose structure is optimized for ground state preparation. For a fair comparison, we transpile the HEA on the basis gateset of \texttt{IBM Torino}.}
\begin{table*}[h!]
\small
\centering
\begin{tabular}{@{}lcccc@{}}
\hline
Method           & Error             & Depth & 1Q gates & 2Q gates \\
\hline
\textbf{GRL (2-gadget)}   & $\mathbf{7.2 \times 10^{-9}}$ & \textbf{13}  & \textbf{12}       & \textbf{9}       \\
QNEAT~\cite{giovagnoli2023qneat}            & $3.2\times10^{-6}$            & 29    & 32       & 14      \\
Hybrid EA~\cite{sunkel2025quantum}            & $2.2\times10^{-6}$            & 15    & 16      & 7      \\
{Const. depth (TFIM square)}~\cite{camps2022algebraic}            & $1.4\times10^{-6}$            & 54   & 81      & 12      \\
{Const. depth (TFIM triangle)}~\cite{camps2022algebraic}            & $1.9\times10^{-6}$            & 51   & 73      & 10      \\
HEA (2 layers)~\cite{kandala2017hardware}   & $8.5\times10^{-9}$ &    39   & 49       & 12      \\
HEA (3 layers)   & $4.5\times10^{-8}$ &   63    & 74       & 19      \\
HEA (4 layers)   & $4.5\times10^{-8}$ &   81    & 105      & 27      \\
\hline
\end{tabular}
\caption{{{Comparison of gadget reinforcement learning (GRL) with non-RL baselines for preparing the ground state of 3-qubit TFIM}. For hardware-efficient ansatz (HEA), each layer contains \texttt{RY} gate followed by \texttt{CX} gates connected in cyclic manner. For a fair comparison we transpile the HEA {and the constant depth (const. depth) TFIM square/triangle ansatzes} on the basis gateset of \texttt{IBM Torino}. The results show that GRL consistently achieves lower error and circuit resource usage compared to non-RL baselines.}}
\label{tab:grl-vs-nonrl-bench}
\end{table*}

{Table~\ref{tab:grl-vs-nonrl-bench} summarizes the key metrics for optimizing the 3-qubit TFIM ground state. This shows that GRL (with bold in the table) substantially outperforms QNEAT, constant depth TFIM ansatz and the HEA, yielding circuits with significantly lower error, reduced gate counts, and more balanced resource composition.} We observe that GRL substantially outperforms both QNEAT and HEA, yielding circuits with significantly lower error, reduced gate counts, and more balanced resource composition. A detailed ablation study of GRL under noise and noiseless cases is provided in Supplementary Note 8.2 in Supplementary Information. Moreover, in Supplementary Note 6 in Supplementary Information, we further elaborate on the performance of GRL for $N=3$ qubit TFIM. While one gadget achieves an error of $10^{-4}$ for $N=3$, adding a second gadget significantly enhances performance, enabling machine precision in both easy and hard regimes. In contrast, the RL-only agent struggles to learn due to the required search depth, particularly in the hard regime ($h=1$), where it produces high-error solutions. Additional gadgets allow RL to find substantially better solutions.

\subsection{Scalability}
The program synthesis approach is computationally feasible for shallow circuits on a CPU. While scaling still suffers from the combinatorial explosion for deep quantum circuits, the gadgets found using smaller and shallower quantum systems can tackle problems up to 6-qubit as we show in the upcoming section. We provide additional details on how scalable program synthesis is outside the GRL framework in Supplementary Note 9 in the Supplementary Information. The investigation demonstrates the scalability of program synthesis up to 5-qubit circuits, with synthesis times ranging from about 1 minute for 2-qubit, depth-5 circuits up to a few hours for 5-qubit, depth-10 circuits, indicating some feasibility for small problem sizes despite the combinatorial growth at larger depths. Gadgets extracted from 2-qubit TFIM circuits generalize effectively to systems up to 10 qubits (which is shown in the later section), i.e., five times larger, highlighting their transferability and efficiency. This suggests that gadgets synthesized for 5-qubit systems could potentially scale to much larger problems, offering a promising direction for future work in improving performance and scalability.

{To assess the scalability of GRL training, we also investigate its transfer capabilities as the TFIM size increases. In Figure~\ref{fig:scalability-of-gadgets}, we demonstrate that gadgets derived from a simple 2-qubit TFIM can be effectively applied to larger TFIMs, achieving improved error rates for finding the ground state in $4$-, $5$-, and $6$-qubit systems. Even with just one gadget, GRL outperforms baseline RL methods, showcasing its efficient scalability by leveraging reusable circuit components.}

{To further emphasize scalability in Supplementary Note 8.4 in the Supplementary Information, we show that the GRL can successfully be extended to a 10-qubit TFIM. This demonstrates the potential for robust performance on increasingly large quantum systems, well beyond previously reported limits.}

\subsection{Found circuits are suitable for real hardware}
\paragraph{More compact circuits for real hardware} We compare PQCs from GRL with state-of-the-art RL methods for finding the TFIM ground state. We benchmark against curriculum reinforcement learning~\cite{patel2024curriculum} using a universal gateset (\texttt{RX}, \texttt{RY}, \texttt{RZ}, \texttt{CX}) as in~\cite{patel2024curriculum,kundu2024reinforcementqas}, comparing it to GRL with an extended action space including gadgets. The GRL action space incorporates the \texttt{IBM Heron} processor's native gateset (in Supplementary Note 12 in the Supplementary Information, we outline the hardware topology) and composite gates derived from top-performing PQCs for 2-qubit TFIM at $h=10^{-3}$ and $h=5\times10^{-2}$. We estimate the ground state of 2-qubit and 3-qubit TFIM at the phase change point ($h=1$). GRL-obtained circuits achieve similar error to RL, but the circuits are more compact when transpiled for real quantum hardware. Supplementary Table~4 summarizes results after transpiling in \texttt{IBMQ Torino} (part of \texttt{IBM Heron} processor). Furthermore, in Supplementary Note 8.3 in the Supplementary Information, we show that the GRL uses ~$3\times$ fewer \texttt{CZ}, \texttt{RZ}, and \texttt{SX} gates for similar error (in the order of $\sim10^{-4}$) in 3-qubit TFIM when transpiled on quantum hardware. This suggests an advantage in solving problems directly with GRL, consisting of target hardware components and gadgets, rather than first finding solutions in a universal gateset and then transpiling for the target hardware.

\paragraph{Improved performance on real hardware} 
Here we compare the length and the performance of the circuits obtained to solve the $2$ and $3$-qubit TFIM ground state at the phase change point ($h=1$) using the RL agent with a universal gateset (i.e. \texttt{RX}, \texttt{RY}, \texttt{RZ} and \texttt{CX}) and GRL agent with an extended action space consisting of gateset of the {IBM Heron} processor and one additional {gadget}. The performance of the GRL and RL is summarized in Supplementary Table~4. {Here, for the 2- and 3-qubit TFIM, we use the circuit with the smallest gate count obtained in the noiseless setting, whose results are elaborated in Supplementary Note 8.3 in the Supplementary Information. Specifically, in the noisy simulation, the optimal 2-qubit TFIM circuit comprises 3 2-qubit gates and 6 1-qubit gates, while the 3-qubit circuit includes 5 2-qubit gates and 30 1-qubit gates. The results show consistent improved performance with GRL across multiple backends compared with simple RL. The illustration of the transpiled circuits in real quantum hardware is provided in Supplementary Note 8.3 in the Supplementary Information.}

{To examine how the number of 2-qubit gates and overall circuit depth affect GRL performance under noise, we analyze 3-qubit TFIM circuits trained in an ideal (noiseless) environment and subsequently test them using the \texttt{IBM Torino} fake backend. For three representative GRL-learned circuits, we obtain ground-state energies of $-3.366$ (5 CZ and 30 1-qubit gates), $-3.312$ (7 CZ and 54 1-qubit gates), and $-3.335$ (11 CZ and 27 1-qubit) on the backend. Circuits with a higher number of CZ generally exhibit increased error compared to those with fewer CZ gates, even when the overall depth is comparable or slightly reduced. Although certain learned circuits achieve shallower depths, their elevated number of CZ per layer can still degrade accuracy in noisy conditions. This observation is consistent with known limitations of both NISQ and fault-tolerant settings~\cite{niu2022parallel,zhu2024ecmas}, where extensive 2-qubit parallelism complicates quantum error correction. These findings highlight that, while GRL can discover compact circuit architectures, maintaining low multi-qubit gate counts and shallow depths remains crucial for reliable performance on real hardware. Future progress in this direction may come from improved reward function design or reinforcement learning state representations~\cite{zen2025quantum} that explicitly balance circuit compactness, multi-qubit gate usage, and target accuracy. Developing such balanced formulations remains an open research challenge.}

A discussion on how the individual gadgets are impacted by noise is provided in Supplementary Note 11 in the Supplementary Information.

{\subsection{Generalization Beyond TFIM}
To address the generalizability of GRL, we tackle the problem of finding the ground state of the 3-qubit H$_2$ molecule (3$-$H$_2$). In this experiment, the \texttt{CZ$_{ij}$RZ$_i$CZ$_{ij}$} (where i and j are qubit indices) and \texttt{X$_i$CZ$_{ij}$} gadget, which are originally learned from the 2-qubit H$_2$ problem, were transferred and applied to the larger 3-qubit TFIM. We observe, GRL with the transferred \texttt{CZ$_{ij}$RZ$_i$CZ$_{ij}$} gadget achieves a lower asymptotic error threshold than standard RL, representing a relative improvement of approximately 9.63\%. Moreover, enabling the transfer and combination of both the \texttt{CZ$_{ij}$RZ$_i$CZ$_{ij}$} and \texttt{X$_i$CZ$_{ij}$} gadgets leads to a further reduction in error, with the GRL approach achieving an improvement of about 43.5\% over the RL baseline. This preliminarily solidifies the application of GRL to quantum chemistry problems. Further investigation on how the GRL performs for more complicated molecules remains an open direction for future extension of this research.
The variation of error threshold for 3-H$_2$ for GRL with 1- and 2-gadgets is illustrated in Supplementary Note 8.5 in the Supplementary Information.

\section{Discussion}
\label{sec:outlook}
In this paper, we have shown how to learn reusable components from different regimes for efficiently building quantum circuits that solve some given problems.
Instead of considering a single specific problem, we start from a trivial regime and gradually tackle the harder one.
By finding the ground state in the low transverse field regime, we discover sequences of recurrent gates, and we can extract them as gadgets and use them to extend the action space of subsequent iterations.
This proves to be very effective because it largely reduces the required depth of the circuit at the cost of a slightly increased breadth of the search.
In other words, the extracted gates serve as a data-driven inductive bias for solving the given class of problems.

In terms of shortcomings of our approach, the main overhead to consider is the necessity of performing multiple iterations.
In particular, it is important that the target class of problems has a structure with different degrees of difficulty: if the problem is too difficult, the reinforcement learning agent does not receive any signal, it will only learn to produce random circuits, and the extracted gates will not necessarily be useful. On the other hand, if one regime is trivial and the other one is too hard, there is a low chance of generalization.
A strict separation of regimes is not necessary, though, as long as we provide batches of problems that include different ranges of difficulty, our algorithm may be able to bootstrap.
Also, to extend the actions of the reinforcement learning agent, multiple approaches are possible. In our example, we reinitialized the agent after extending the action space. However, smarter approaches, for example, by just adding extra output neurons at the last layer of the policy, associating them with the added gadgets, may allow starting from the previous policy while adding a small bias to encourage the exploration of the new action.

Our technique is general and can be extended to other quantum problems. For instance, we can efficiently solve challenging correlated quantum chemistry instances by leveraging gadgets from simpler ones. Easy instances involve smaller action spaces, and as the action space grows or the accuracy requirements for the ground state increase, the problem becomes more difficult~\cite{mccaskey2019quantum,de2023complete}. 
{We have shown a promising example application to a 3-qubit H$_2$ molecule. While our findings provide preliminary evidence for the application of GRL beyond TFIM, a deeper investigation is required for more complex molecular systems, which we leave as an important direction for future work.}
This approach can also be applied to quantum optimization, simulation, and machine learning, where easy instances help address more complex scenarios.
Furthermore, it may be suitable for real hardware optimizations. Indeed, it allows you to explicitly define the elementary gates to use for the decomposition, as opposed to finding the solution in a high-level gate set first (e.g. rotation gates $\texttt{RX}$, $\texttt{RY}$, and $\texttt{RZ}$) and transpiling them later.
This can arguably produce more efficient circuits.
Also, penalties for the length of the circuit or the use of specific gates could be enforced, encouraging gates that are more reliable or cheaper to implement on real hardware.
In addition, the elementary components could also be modified to include some model of the noise on the real hardware, thus possibly finding a solution for some quantum problem that already includes some noise mitigation effects. The limitations and future directions are outlined in Supplementary Note 1 in the Supplementary Information. 

\section{Data availability}
The dataset generated by running GRL and RL throughout the paper is available upon request from the corresponding author.

\section{Code availability}
{The GRL framework is available on} \textcolor{blue}{GitHub} at \url{https://github.com/Aqasch/Gadget_RL}. 


\section*{Acknowledgements}
A.K. acknowledges funding from the Research Council of Finland through the Finnish Quantum Flagship project 358878 (UH). The authors wish to thank the Finnish Computing Competence Infrastructure (FCCI) for supporting this project with computational and data storage resources. L.S. acknowledges that parts of the computations in this work were run at facilities supported by the Scientific Computing Core at the Flatiron Institute. Work at the Flatiron Institute is supported by the Simons Foundation.

\section*{Author contributions}
A.K. developed the concept of the study. L.S proposed and
developed the application of program synthesis in the quantum domain. A.K. and L.S. implemented the gadget reinforcement learning framework. A.K. proposes the TFIM and quantum chemistry application and trains the reinforcement learning agent. L.S. synthesizes the best quantum circuits for the GRL framework. Both authors interpreted the results, finalized, reviewed and revised the manuscript.

\section*{Competing interests}
The authors declare no competing interests.

\bibliography{bibliography}
\bibliographystyle{unsrt}

\appendix
\onecolumn

\section*{Supplementary Note 1: Limitations and future work}\label{appndx:limitations_future_work}
Despite the demonstrated advantages of the gadget reinforcement learning (GRL) framework, several limitations and avenues for future work remain:
\begin{enumerate}
    \item {Classical gadget discovery:} While our approach focuses on quantum circuit gadgets, the GRL framework is general and could be adapted to classical environments. Investigating the discovery and utility of ``classical gadgets''---reusable composite operations---in classical reinforcement learning or optimization settings is an intriguing direction for future research.
    
    \item {Noise modelling and robustness:} Our experiments employ a simple probabilistic single-qubit noise model for gadgets. However, real quantum hardware exhibits a broader range of noise processes. A thorough investigation of GRL's robustness under more realistic, device-specific noise models and the development of noise-aware gadgets for a broader class of problems is an important open challenge.
    
    \item {Transferability of gadgets:} We only demonstrate that gadgets learned from 2-qubit TFIM instances can improve performance for up to 6-qubit TFIM systems. It remains unclear how well these gadgets generalize when the underlying problem class or Hamiltonian structure is changed. Systematic studies are needed to assess the universality and limitations of gadget transfer across problem domains and circuit sizes.
    
    \item {Computational overhead and scalability:} The iterative nature of GRL, particularly the program synthesis step, introduces additional computational overhead. Although we show feasibility for shallow circuits and up to 6 qubits, scaling to deeper circuits and larger systems remains a challenge. Future work should explore more efficient synthesis algorithms and strategies for incremental action space expansion without full agent retraining.

    \item {Extension to other quantum problems:} This work focuses exclusively on the transverse field Ising model (TFIM). We have not addressed other important quantum problems, such as quantum chemistry Hamiltonians or those arising in condensed matter physics. Extending GRL to these domains is a promising direction for future research.

    \item {{Extending the current benchmark:} As a direction for future work, gadget reinforcement learning (GRL) could be systematically compared to established search-based quantum circuit synthesis algorithms such as SQUANDER~\cite{rakyta2022approaching, rakyta2022efficient} and QSearch~\cite{davis2020towards} to further quantify its practical benefits. Such studies would help clarify GRL’s advantages and limitations relative to direct ansatz construction for structured problems like TFIM.}
    
\end{enumerate}
These limitations highlight the need for further research to broaden the applicability, robustness, and efficiency of the GRL framework in both quantum and classical domains.

\section*{Supplementary Note 2: Feedback-driven curriculum reinforcement learning~\cite{patel2024curriculum}}
\label{app:feedback_driven_crl}
In this section, we review feedback-driven curriculum reinforcement learning.
During learning, the agent maintains a pre-defined threshold \(\zeta_2\) representing the lowest energy observed so far, updating it based on defined rules. Initially, \(\zeta_2\) is set to a hyperparameter \(\zeta_1\). When a lower threshold is found, \(\zeta_2\) is updated to this new value. A {fake minimum energy} hyperparameter, \(\mu\), serves as a target energy, approximated by the following:
\begin{equation}
    \text{fake minimum energy} = (N-1)\times(-J)+ N\times(-h),
\end{equation}
where $N$ is the number of interacting spins, $J$ is the coupling strength between the spins and $h$ is the strength of the magnetic field.

Without amortization, the threshold updates to \(|\mu - \zeta_2|\) when \(\zeta_2\) changes; with amortization, it becomes \(|\mu - \zeta_2| + \delta\), where \(\delta\) is an amortization hyperparameter. The agent then explores subsequent actions and records successes.

Two threshold adjustment rules apply: a greedy shift to \(|\mu - \zeta_2|\) after \(G\) episodes (where \(G\) is a hyperparameter) and a gradual decrease by \(\delta / \kappa\) with each successful episode, where \(\kappa\) is a shift radius hyperparameter. If repeated failures occur after setting the threshold to \(|\mu - \zeta_2|\), it reverts to \(|\mu - \zeta_2| + \delta\), allowing the agent to backtrack if stuck in a local minimum.

\section*{Supplementary Note 3: Technical details of the library building algorithm}
\label{app:library-algorithm}

The library building algorithm analyzes a set of circuits $\mathcal D$ in relation to a given set of elementary gates $g$, called grammar.
In this framework, each grammar \( g \) consists of elementary gates (also called ``primitives'')  with assigned probabilities based on usage frequency in the dataset.
When we consider whether we should add a new gadget to our set of elementary gates, we compare the grammar with and without the new gadget, and accept the new gate if we improve the grammar score.
This quantity evaluates how good a grammar is to represent the given dataset $\mathcal{D}$, trading off the likelihood of sampling circuits from the dataset with an approximation of the complexity of the grammar itself.
In particular, given a set of circuits \( \mathcal{D} \), we define the grammar score \( S \), representing the grammar’s efficiency in describing the circuits, as

\begin{equation}
    S_\mathcal{D}(g) = L_g(\mathcal{D}) - \lambda |g| - k \sum_{p \in g} |p|,
    \label{eq:grammar_score}
\end{equation}

where:
\begin{itemize}
    \item \( |g| \): the number of components in grammar \( g \),
    \item \( p \): a component or building block in the grammar,
    \item \( |p| \): the number of elementary gates in \( p \),
    \item \( \lambda = 1 \) and \( k = 1 \) are hyperparameters.
\end{itemize}

The first term represents the likelihood of reproducing the observed circuits, while the last two terms are complexity regularizers, inspired by the minimum description length principle. The likelihood \( L_g(\mathcal{D}) \) of a grammar is approximated with the probability of randomly sampling the circuits in the dataset using the grammar gate probabilities. Each circuit is weighted by the accuracy in solving the task (measured as the opposite of the energy).

The main hyperparameters that we consider to tune the algorithm are:
\begin{enumerate}
    \item {Arity}: This controls the maximum number of arguments a component can have, or equivalently, the maximum number of qubits an extracted gate can act on. Here, we set \(\text{arity}=2\).
    \item {Pseudocounts}: A constant shift in the usage frequency, which adjusts the log-likelihood estimation by ensuring each component is treated as though it is used at least once, even if unobserved. This allows patterns to be considered useful only if they appear frequently in the dataset. We set \(\text{pseudocounts}=10\).
    \item {Structure Penalty \( k \)}: This regularizes the tradeoff between grammar likelihood and complexity. Lower penalties yield higher likelihoods but may overfit, while higher penalties result in simpler grammars that generalize better. We set \(\text{structurePenalty}=1\).
\end{enumerate}

For a more technical descriptions of the $\lambda$-calculus tree structures and their efficiency, see~\cite{DREAMCODER_Ellis2020DreamCoderGG, DiscoveringQuantumCircuits_Sarra_2024}.

\section*{Supplementary Note 4: Implementation details}
In this section we provide further details on the implementation of the RL-state encoding scheme, the RL-algorithm and the agent hypermeter settings.
\label{app:implementation_details}

\subsection*{Supplementary Note 4.1: Quantum circuit encoding}\label{app:tensor_based_encoding}
We employ a refined version of the tensor-based binary encoding introduced in~\cite{kundu2024kanqas}, which is inspired by the encoding presented in~\cite{patel2024curriculum}, to capture the architecture of a parametric quantum circuit (PQC), specifically by encoding the sequence and arrangement of quantum gates. Unlike the encoding presented in~\cite{patel2024curriculum}, which is only the function of the number of qubits $N$, the refined encoding is a function of $N$ and the number of 1-qubit gates $N_\text{1q}$. This makes it suitable for the encoding of a broad range of action spaces and enables the agent to access a complete description of the circuit. To ensure a consistent input size across varying circuit depths, we construct the tensor for the maximum anticipated circuit depth.

To build this tensor, we define the hyperparameter $T_{\text{max}}$, which restricts the number of allowable gates (actions) across all episodes. A {moment} in a PQC refers to all simultaneously executable gates, corresponding to the circuit's depth. We represent PQCs as three-dimensional tensors where, at the start of each episode, an empty circuit of depth $T_{\text{max}}$ is initialized. This tensor is dimensioned as $\left[T_{\text{max}} \times \left((N+N_\text{1q})\times N\right)\right]$, where \( N \) denotes the number of qubits and \( N_\text{1q} \) the number of 1-qubit gates. Each matrix slice within the tensor contains \( N \) rows that specify control and target qubit locations in \texttt{CNOT} gates, followed by either 3 rows (for \texttt{RX}, \texttt{RY} and \texttt{RZ}) or 3 rows (for \texttt{SX}, \texttt{X}, \texttt{RZ}) to indicate the positions of 1-qubit gates. When we update the action space by incorporating the {gadgets}, (which are the composite gateset found using the program synthesis algorithm) then, depending on the added gadget, we update the size of the tensor. 
After {gadgetizing} the action space, we rerun the RL agent with the extended encoding of the PQCs.

\subsection*{Supplementary Note 4.2: Double Deep Q-Network (DDQN)}\label{appndx:ddqn}
Deep Reinforcement Learning (RL) methods employ Neural Networks (NNs) to refine the agent's policy in order to maximize the cumulative return: \begin{equation} G_t = \sum_{k=0}^{\infty} \gamma^k r_{t+k+1}, \end{equation} where $\gamma \in [0,1)$ denotes the discount factor. An action value function is assigned to each state-action pair $(s,a)$, capturing the expected return when action $a$ is taken in state $s$ at time $t$ under policy $\pi$: \begin{equation} q_\pi (s,a) = \mathbb{E}_\pi [G_t | s_t = s, a_t = a]. \end{equation}

The objective is to find an optimal policy that maximizes the expected return. This can be achieved through the optimal action-value function $q_\ast$, which satisfies the Bellman optimality equation: \begin{equation} q_\ast (s,a) = \mathbb{E} \bigg[ r_{t+1} + \max_{a'} q_\ast (s_{t+1},a') \big| s_t = s, a_t = a \bigg]. \end{equation}

Rather than solving the Bellman equation directly, value-based RL focuses on approximating the optimal action-value function through sampled data. Q-learning, a widely used value-based RL algorithm, initializes with arbitrary Q-values for each $(s,a)$ pair and iteratively updates them to approach $q_\ast$. The update rule for Q-learning is: \begin{eqnarray}
    Q(s_t,a_t) &\leftarrow& Q(s_t,a_t) + \alpha \big( r_{t+1} + \gamma \max_{a'} Q(s_{t+1},a') - Q(s_t,a_t) \big),
\end{eqnarray}
where $\alpha$ is the learning rate, $r_{t+1}$ is the reward received at step $t+1$, and $s_{t+1}$ is the resulting state after taking action $a_t$ in state $s_t$. Convergence to the optimal Q-values is guaranteed under the tabular setup if all state-action pairs are visited infinitely often~\cite{melo2001convergence}. To promote exploration in Q-learning, an $\epsilon$-greedy policy is adopted, defined as:
\begin{align}
    \pi(a|s)\coloneqq \begin{cases}
                          1-\epsilon_t & \text{if $a = \max_{a'} Q(s,a')$}, \\
                          \epsilon_t   & \text{otherwise}.
                      \end{cases}
\end{align}
This $\epsilon$-greedy policy adds randomness during learning, while the policy becomes deterministic after training.

To handle large state and action spaces, NN-based function approximations are used to extend Q-learning. Since NN training relies on independently and identically distributed samples, this requirement is met through experience replay.
With experience replay, transitions are stored and randomly sampled in mini-batches, reducing the correlation between samples.
For stable training, two NNs are employed: a policy network that is frequently updated, and a target network, which is a delayed copy of the policy network.
The target value $Y$ used in updates is given by: \begin{equation} Y_{\text{DQN}} = r_{t+1} + \gamma \max_{a'} Q_{\text{target}}(s_{t+1},a'). \end{equation}

In the double DQN (DDQN) approach, the action used for estimating the target is derived from the policy network, minimizing the overestimation bias observed in standard DQN. The target is thus defined as: \begin{equation} Y_{\text{DDQN}} = r_{t+1} + \gamma Q_{\text{target}} \big( s_{t+1}, \arg \max_{a'} Q_{\text{policy}}(s_{t+1},a') \big). \end{equation} This target value is then approximated through a loss function, which in our work is chosen to be the smooth L1-norm given by
\begin{equation}
    \text{SmoothL1}(x) =
    \begin{cases}
        0.5 x^2   & \text{if } |x| < 1, \\
        |x| - 0.5 & \text{otherwise}.
    \end{cases}
\end{equation}

\subsection*{Supplementary Note 4.3: Reinforcement learning agent hyperparameters}\label{app:hyperparams_grl_crl}
The hyperparameters of the double deep-Q network algorithm were selected through coarse-grain search, and the employed network architecture depicts a feed-forward neural network whose hyperparameters are provided in supplementary tab.~\ref{tab:agent_hyper}.
\begin{table}[h!]
\centering
\small
\begin{tabular}{l c || l c}
{Parameter} & {Value} & {Parameter} & {Value} \\ \hline
Batch size & 1000 & Network optimizer & Adam~\cite{kingma2014adam} \\
Memory size & 20000 & Learning rate & $10^{-4}$ \\
Neurons & 1000 & Update target network & 500 \\
Hidden layers & 5 & Final gamma & $5\times10^{-3}$ \\
Minimum epsilon & $5\times10^{-2}$ & Epsilon decay & 0.99995 \\
\end{tabular}
\caption{GRL and RL agent hyperparameters.}
\label{tab:agent_hyper}
\end{table}

In the implemented agents, we greedily update the threshold ($\zeta$) after 2000 episodes, with an amortization radius set at $10^{-4}$.
This amortization radius decreased by $10^{-5}$ after every $50$ successfully solved episode,
beginning from an initial threshold value of $\zeta_1 = 5\times10^{-3}$. Moreover, in each episode, we set the total number of steps $T_\text{max} = 20$ for 2-qubit TFIM and $T_\text{max} = 50$ for 3-qubit TFIM.

Throughout this paper, we utilize a gradient-free COBYLA optimizer~\cite{powell1994direct} with hyperparameter settings similar to ref.~\cite{virtanen2020scipy} and $1000$ iterations at each step of an episode to optimize the PQCs.

\section*{Supplementary Note 5: The transverse field Ising model in different regimes}\label{ref:degeneracy_TFIM}

As shown in supplementary fig.~\ref{fig:TFIM_hardness}, by looking at the ground state energy gap, we can identify three different regimes in the Transverse Field Ising Model:
\begin{enumerate}
    \item In the low external field, the first excited state is almost degenerate with the ground state, therefore, it is easy to find a low-energy state.
    \item the regime where $h\simeq 0.1$, where the energy gap increases and the ground state starts to have a visibly different energy from the first excited state;
    \item $h\gg 0.1$ where the energy gap is larger and the ground state energy is much smaller than that of the first excited state.\todo{but we also said that the problem becomes easy again for very large $h$.}
\end{enumerate}
\begin{figure}[h!]
    \vskip 0.2in
    \begin{center}
        \centerline{\includegraphics[width=0.5\columnwidth]{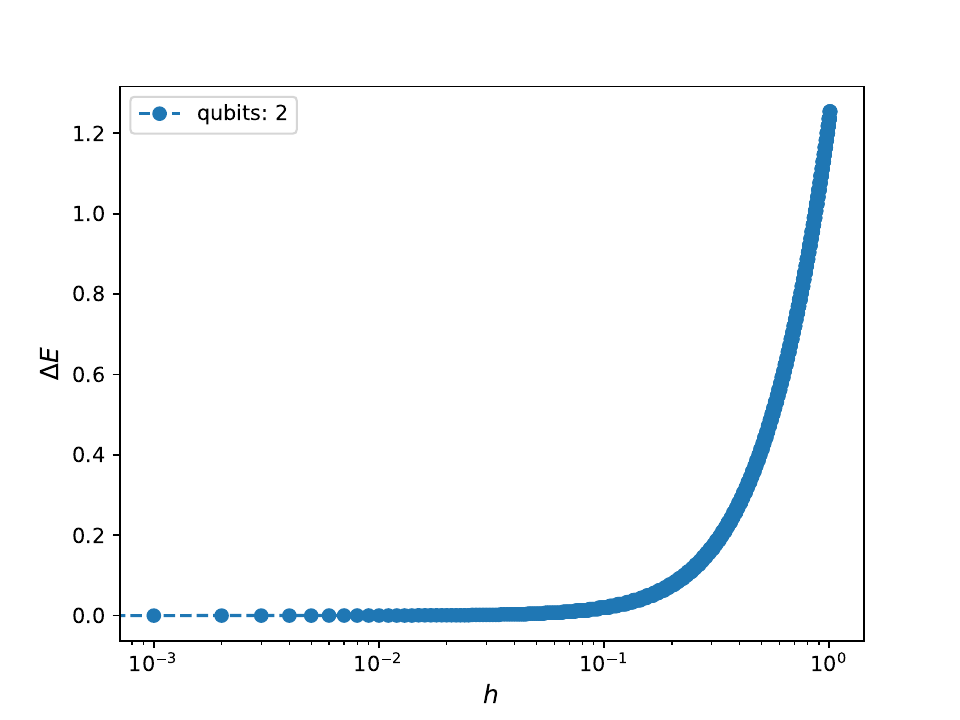}}
        \caption{Energy gap between the first excited state and the ground state of the TFIM model as a function of the transverse field strength. The separation $\Delta E$ is negligible till $h=10^{-1}$. Hence, due to energy degeneracy, it is easy to find a good energy approximation for $h\leq10^{-1}$. The problem becomes harder when we choose, $h\geq10^{-1}$ as $\Delta E$ becomes non-negligible.}
        \label{fig:TFIM_hardness}
    \end{center}
    \vskip -0.2in
\end{figure}
In our experiments, we choose values of $h$ in the three different regimes, so that we first try to tackle the easier case ($h=0.001$, small energy gap, almost degenerate), and use the extracted gadgets to solve the harder cases, $h=0.5$ (larger energy gap, as shown in \ref{fig:TFIM_hardness}) and $h=1$.

\section*{Supplementary Note 6: GRL on 2- and 3-qubit TFIM}\label{app:rewards_return_3qubit}

Supplementary fig.~\ref{fig:N=3_reward} illustrates the cumulative performance of RL and GRL agents over a series of episodes for solving the 3-qubit transverse field Ising model (TFIM). Key metrics, such as error threshold, rewards, and returns, are plotted against the number of episodes to evaluate the effectiveness of each approach.
\begin{figure}[h!]
    \centering
    \includegraphics[width=\linewidth]{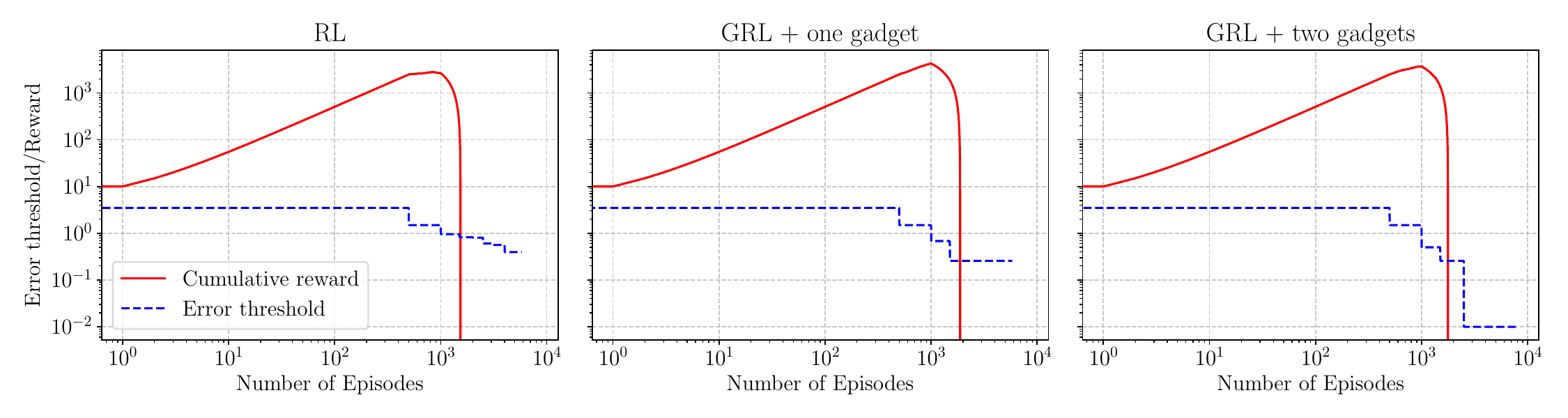}
    \caption{{Comparative performance of RL and GRL agents in solving the 3-qubit TFIM}. The plots show cumulative rewards, error scaling, and success rates over episodes for RL-only, GRL with one gadget, and GRL with two gadgets. GRL agents demonstrate improved stability, faster convergence, and higher success rates, particularly when multiple gadgets are incorporated}
    \label{fig:N=3_reward}
\end{figure}

The RL-only agent struggles to achieve stable rewards across episodes, showing significant fluctuations and slow convergence. In contrast, GRL agents exhibit steady improvements in cumulative rewards, particularly when gadgets are incorporated into the action space.

For GRL with one gadget, the error decreases significantly in early episodes but plateaus at a higher value, indicating limited capability to reach machine precision. 
GRL with two gadgets further reduces the error, demonstrating the benefits of adding more extracted gadgets for addressing complex regimes.
The frequency of successful episodes (defined by achieving the ground state approximation within a predefined threshold) increases with the number of gadgets in the GRL setup. 
The RL-only agent rarely achieves success, particularly in the harder regimes.

These results highlight the scalability and robustness of GRL agents. 
Incorporating gadgets not only accelerates the learning process but also enables the agent to achieve lower error thresholds and higher success rates, even for challenging configurations like the 3-qubit TFIM.
\begin{figure}[h!]
    \centering
   \includegraphics[width=0.8\linewidth]{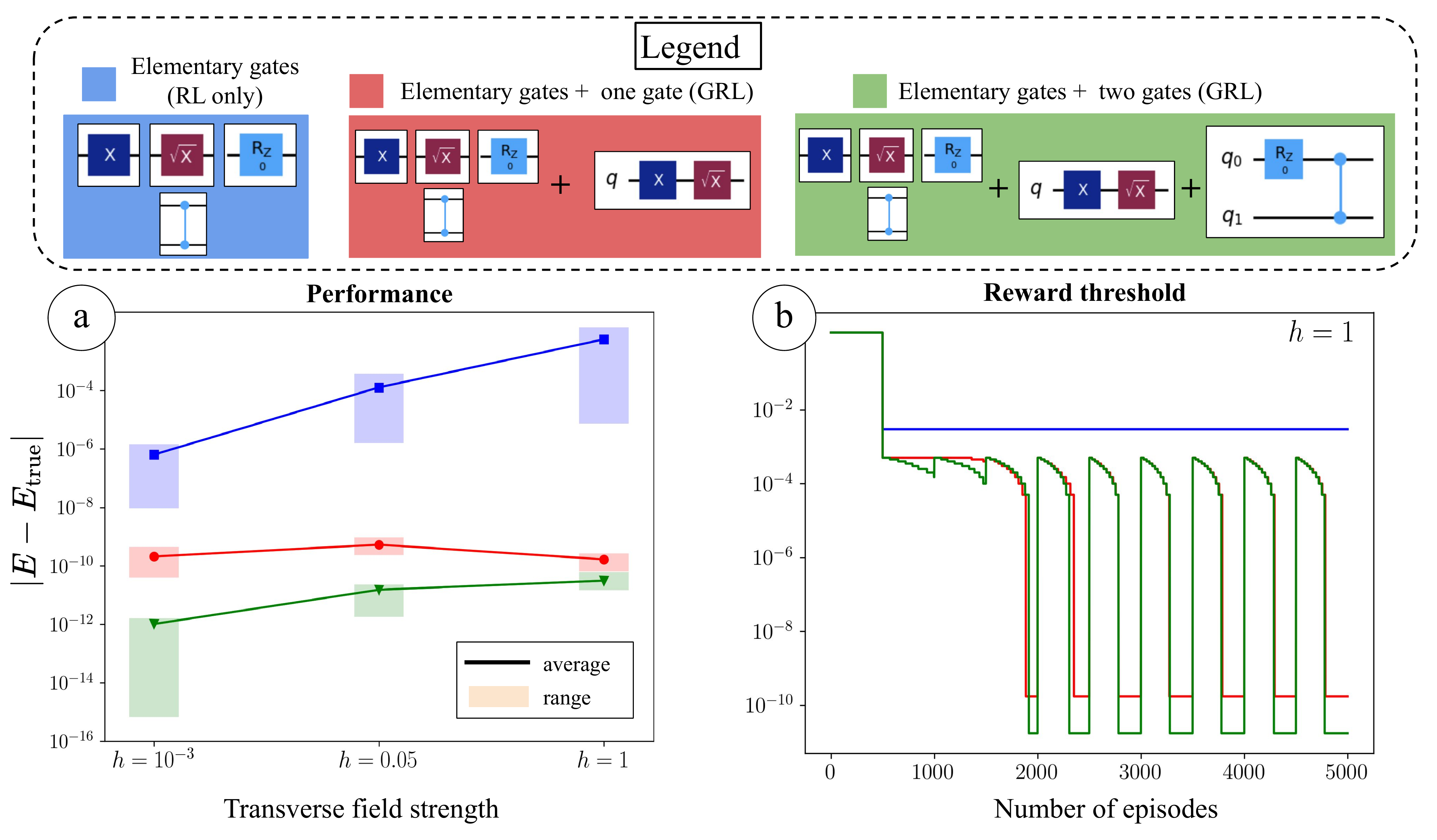}%

    \caption{\small{{Results for the 2-qubit transverse field Ising model (TFIM)}.
We compare reinforcement learning-only (blue) with gadget reinforcement learning (GRL) using 1-gadget (red) and 2-gadget (green), as given in the legend.
(a) Compares error scaling with varying transverse field strength under a fixed compute budget (a max of 48-hour GPU run). Solid lines show averages over multiple runs; shaded areas indicate solution ranges (smallest values are most relevant). GRL with 1-gadget (red) and 2-gadget (green) achieves high accuracy for $h=1$ compared to RL-only (blue).
(b) plots RL reward thresholds during training for $h=1$, showing GRL finds circuits with lower cost. Without gadget extraction, accuracy is limited to $10^{-3}$, while GRL achieves machine precision. A similar illustration for 3-qubit TFIM is shown in Supplementary Note 7 in Supplementary information}.}
    \label{fig:2_qubit_results}%
\end{figure}

\begin{figure}[t!]
    \centering
   \includegraphics[width=0.8\linewidth]{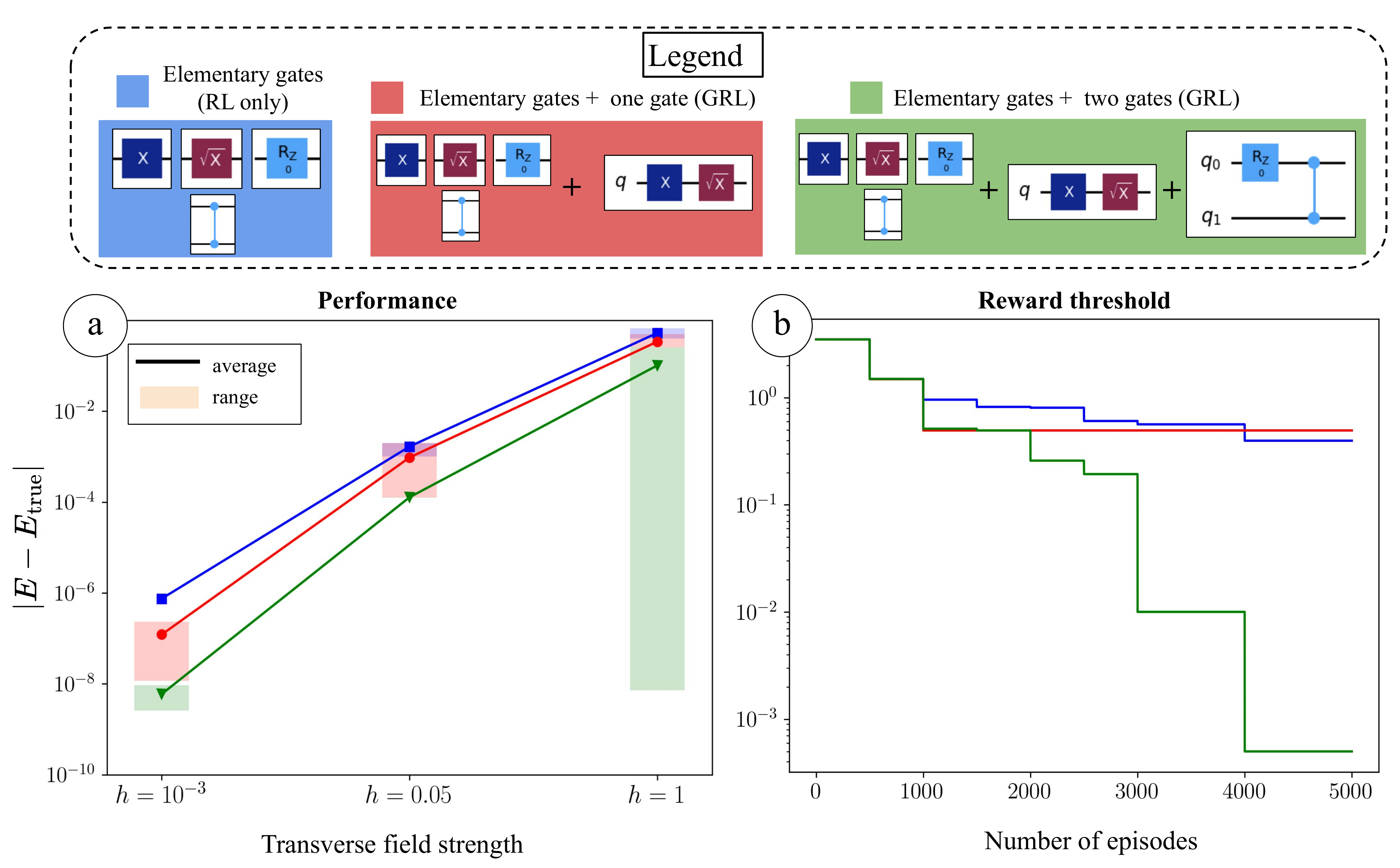}%

    \caption{ {\small{GRL with two gadgets (green) overcomes the training bottleneck of RL-only (blue) and GRL with one gadget (red).}
The subplots (a) and (b) compare error scaling and RL reward thresholds. Under a fixed compute budget and varying transverse field strength, RL and GRL with one gadget achieve accuracies around $0.1$, whereas GRL with two gadgets attains machine precision. It should be noted that the gadgets that are used in this simulation are the same ones extracted while solving the $N=2$ qubit TFIM.}
 }
    \label{fig:3_qubit_results}%
\end{figure}

In supplementary fig.~\ref{fig:3_qubit_results} and supplementary fig.~\ref{fig:2_qubit_results}, we show that as we add more gadgets in the GRL framework, the error in the preparation of the ground state reduces, making the gadgets found in for 2-qubit program synthesis reusable.
The inclusion of additional gadgets systematically reduces errors and increases the frequency of successful episodes, making this approach a viable solution for tackling more complex quantum systems. This analysis demonstrates the potential of gadget-based reinforcement learning to enhance agent performance.

\section{Supplementary Note 7: Performance evaluation of GRL}\label{app:numerical-results}
{
Here we present a comprehensive evaluation of our Gadget Reinforcement Learning (GRL) approach for finding the ground state of the Transverse Field Ising Model (TFIM).
Supplementary tab.~\ref{tab:ablation_study_in_details} shows the results on both 2-qubit and 3-qubit TFIM systems across the three distinct regimes: low (\(h=10^{-3}\)), intermediate (\(h=5\times10^{-2}\)), and strong (\(h=1\)) transverse fields. 
In many cases, GRL produces shorter circuits with fewer gates. For example, in the 3-qubit TFIM with \(h=5\times10^{-2}\), 1-gate GRL finds a solution with only 11 gates and 6 depth, compared to 12 gates and 9 depth for RL-only. 
The advantage of GRL becomes more pronounced as we move from 2-qubit to 3-qubit systems, suggesting better scalability compared to RL-only. 
The performance gap between GRL and RL-only varies across different field strengths, with GRL showing particular strength in the low and intermediate field regimes.}
\begin{figure}[h!]
    \centering
    \includegraphics[width=\linewidth]{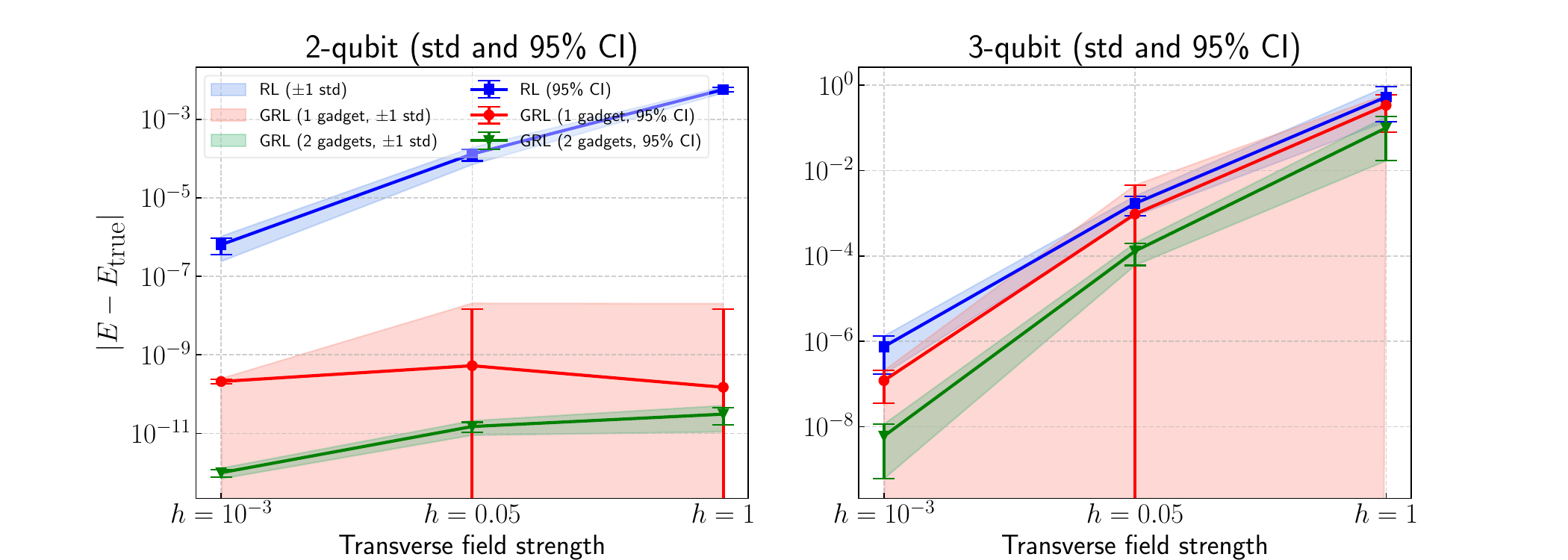}
    \caption{{Illustration of the 95\% confidence intervals and standard deviations for the performance of 2- and 3-qubit TFIM, evaluated across 5 different runs.}}
    \label{fig:confidence}
\end{figure}

{Importantly, we {explicitly report both the 95\% confidence intervals and standard deviations} for all performance measures, and clearly state the number of random seeds used in our evaluation. As illustrated in supplementary fig.~\ref{fig:confidence}, each data point is accompanied by error bars representing both the standard deviation (shaded regions) and the 95\% confidence interval (solid error bars), computed over 5 independent runs with different random seeds. This confirms the consistency and significance of the observed performance differences: by quantifying the statistical uncertainty and inherent randomness, our analysis ensures that the conclusions drawn regarding the advantages of GRL are robust and statistically substantiated.}
\begin{table*}[t!]
    \centering
    \small
    \setlength{\tabcolsep}{4pt}
    \renewcommand{\arraystretch}{1.2}
    \begin{tabular}{@{}c|cc|cccc|cccc@{}}
        Settings                              & Problem                             & \makecell{Field                                                                                                         \\ str.} & \makecell{Avg. \\ err.} & \makecell{Avg. \\ gate} & \makecell{Avg. \\ 2q gate} & \makecell{Avg. \\ depth} & \makecell{Min. \\ err.} & \makecell{Min. \\ gate} & \makecell{Min. \\ 2q gate} & \makecell{Min. \\ depth}\\
        \hline
        \multirow{6}{4.6em}{\makecell{\textbf{2-gate}                                                                                                                                                         \\\textbf{GRL} }} & \multirow{3}{4em}{\makecell{2-qubit\\TFIM}} & h=$10^{-3}$ & $\mathbf{1.0\times10^{-12}}$ & 25.33 & 8.0 & 22.33 & $\mathbf{6.67\times10^{-16}}$ & 21 & 3 & 19 \\
                                              &                                     & h=$5\times10^{-2}$ & $\mathbf{1.5\times10^{-11}}$ & 18.0  & 3.67  & 15.0  & $\mathbf{1.8\times10^{-12}}$ & 8  & 2  & 7  \\
                                              &                                     & h=1                & $3.1\times10^{-11}$          & 13.67 & 3.33  & 10.0  & $1.4\times10^{-11}$          & 9  & 3  & 7  \\
        \cline{2-11}
                                              & \multirow{3}{4em}{\makecell{3-qubit                                                                                                                           \\TFIM}} & h=$10^{-3}$ & $\mathbf{6\times10^{-9}}$ & 20.5 & 2.5 & 12.0 & $\mathbf{2.6\times10^{-9}}$ & 19 & 1 & 12 \\
                                              &                                     & h=$5\times10^{-2}$ & $1.3\times10^{-4}$           & 42.0  & 11.67 & 29.0  & $1.3\times10^{-4}$           & 33 & 3  & 19 \\
                                              &                                     & h=1                & {0.10}                & 41.0  & 8.33  & 28.0  & $\mathbf{7.2\times10^{-9}}$  & 35 & 5  & 25 \\
        \hline\hline
        \multirow{6}{4.6em}{\makecell{\textbf{1-gate}                                                                                                                                                         \\\textbf{GRL} }} & \multirow{3}{4em}{\makecell{2-qubit\\TFIM}} & h=$10^{-3}$ & $2.1\times10^{-10}$ & 18.33 & 5.33 & 14.67 & $3.9\times10^{-11}$ & 8 & 2 & 6 \\
                                              &                                     & h=$5\times10^{-2}$ & $5.3\times10^{-10}$          & 14.33 & 3.33  & 11.0  & $2.3\times10^{-10}$          & 11 & 2  & 9  \\
                                              &                                     & h=1                & $1.5\times10^{-10}$          & 11.67 & 1.67  & 8.0   & $6.6\times10^{-11}$          & 8  & 1  & 6  \\
        \cline{2-11}
                                              & \multirow{3}{4em}{\makecell{3-qubit                                                                                                                           \\TFIM}} & h=$10^{-3}$ & $1.2\times10^{-7}$ & 43.0 & 11.5 & 28.5 & $1.2\times10^{-8}$ & 38 & 6 & 26 \\
                                              &                                     & h=$5\times10^{-2}$ & $\mathbf{9.6\times10^{-4}}$  & 16.0  & 3.67  & 10.33 & $\mathbf{1.3\times10^{-4}}$  & 11 & 2  & 6  \\
                                              &                                     & h=1                & 0.34                         & 40.67 & 25.67 & 31.0  & 0.26                         & 36 & 19 & 27 \\
        \hline\hline
        \multirow{6}{3.2em}{\textbf{RL only}} & \multirow{3}{4em}{\makecell{2-qubit                                                                                                                           \\TFIM}} & h=$10^{-3}$ & $6.4\times10^{-7}$ & 14.33 & 3.0 & 11.33 & $9.5\times10^{-9}$ & 11 & 2 & 9 \\
                                              &                                     & h=$5\times10^{-2}$ & $1.3\times10^{-4}$           & 21.67 & 5.33  & 16.33 & $1.6\times10^{-6}$           & 21 & 3  & 15 \\
                                              &                                     & h=1                & $5.7\times10^{-3}$           & 20.33 & 3.0   & 13.67 & $7.4\times10^{-6}$           & 14 & 2  & 10 \\
        \cline{2-11}
                                              & \multirow{3}{4em}{\makecell{3-qubit                                                                                                                           \\TFIM}} & h=$10^{-3}$ & $7.5\times10^{-7}$ & 18.0 & 9.5 & 13.5 & $7.5\times10^{-7}$ & 11 & 3 & 6 \\
                                              &                                     & h=$5\times10^{-2}$ & $1.7\times10^{-3}$           & 15.67 & 7.0   & 11.67 & $1.0\times10^{-3}$           & 12 & 3  & 9  \\
                                              &                                     & h=1                & 0.53                         & 36.0  & 7.0   & 24.3  & 0.39                         & 29 & 2  & 18 \\
    \end{tabular}
    \caption{\small {Gadget reinforcement learning (GRL) produces better approximations and sometimes even shorter circuits than RL only, especially in the hardest regimes}. Results are obtained by tackling the task of finding the ground state of the transverse field Ising model (TFIM) for 2- and 3-qubits in three different regimes (low, intermediate and strong transverse field). We compare the performance with one and two extracted gadgets and RL only. The average is taken over $5$ different initializations of the neural network, and the minimum is the best-performing instance.}
    \label{tab:ablation_study_in_details}
\end{table*}

{\section*{Supplementary Note 8: Ablation study}}\label{appendx:gadget_transfer_evaluation}

\subsection*{Supplementary Note 8.2: GRL with 1- and 2-gadgets in noiseless case}supplementary tab.~\ref{tab:comparison} presents a comparative analysis of the GRL and RL methods across various qubit configurations, evaluating their performance in terms of error rate, gate count, and circuit depth. 
\begin{table}[h!]
\small
    \centering
    \caption{{Comparison of GRL and RL methods for different qubit configurations to sole the TFIM ground state estimation.}}
    \label{tab:comparison}
    \begin{tabular}{@{}cccccc@{}}
        Qubit & Method & Error & Gates & Depth \\
        \midrule
        3 & \textbf{GRL} & $\mathbf{7.2 \times 10^{-4}}$ & $\mathbf{12}$ & $\mathbf{3}$ \\
         & RL  & $2.1 \times 10^{-3}$ & $15$ & $5$ \\
         \hline
        4 & \textbf{GRL} & $\mathbf{1.5 \times 10^{-3}}$ & $\mathbf{8}$  & $\mathbf{7}$ \\
         & RL  & $2.6 \times 10^{-3}$ & $12$ & $26$ \\
         \hline
        5 & \textbf{GRL} & $\mathbf{3.1 \times 10^{-3}}$ & $\mathbf{18}$  & $\mathbf{5}$ \\
         & RL  & $4.3 \times 10^{-3}$ & $33$ & $25$ \\
         \hline
        6 & GRL & $\mathbf{3.7 \times 10^{-3}}$ & $\mathbf{14}$ & $\mathbf{8}$ \\
         & RL  & $5.0 \times 10^{-3}$ & $40$ & $25$ \\
    \end{tabular}
\end{table}
The results consistently demonstrate that the GRL method outperforms the RL baseline for all tested qubit numbers. For instance, with three qubits, GRL achieves a significantly lower error ($7.2 \times 10^{-4}$) compared to RL ($2.1 \times 10^{-3}$), while also requiring fewer gates and a shallower circuit depth. This trend persists as the number of qubits increases: for 4-, 5-, and 6-qubit, GRL maintains lower error rates and uses fewer gates and reduced depths than RL. The depth advantage of GRL is especially pronounced in the four-qubit scenario, where it achieves a depth of 7 versus RL's 26. These results highlight the scalability and efficiency of the GRL approach, suggesting its strong potential for more complex quantum circuit synthesis tasks.

\subsection*{Supplementary Note 8.3: Performance comparison of transpiled circuits on QPU using RL and GRL}
\label{app:transpiled_circuits}

\begin{table}[t!]
\small
\centering
\setlength{\tabcolsep}{4pt}
\renewcommand{\arraystretch}{1.2}

\begin{subtable}[t]{0.55\textwidth}
\centering
\begin{tabular}{@{}l|cccccc@{}}
\hline
Problem & Method & Metric & \#CZ & \#RZ & \#SX & \#X \\
\hline
\multirow{4}{4em}{\makecell{2-qubit\\TFIM}} 
  & \textbf{GRL} & Average & \textbf{2.0} & \textbf{6.0} & \textbf{4.43} & 2.0 \\
  &  & Minimum & 2 & 5 & 4 & 1 \\
\cline{2-7}
  & RL & Average & 2.0 & 8.08 & 6.62 & 1.75 \\
  & & Minimum & 1 & 6 & 4 & 1 \\
\hline
\multirow{4}{4em}{\makecell{3-qubit\\TFIM}} 
  & \textbf{GRL} & Average & \textbf{6.0} & \textbf{11.0} & \textbf{11.0} & 1.0 \\
  &  & Minimum & 2 & 9 & 7 & 1 \\
\cline{2-7}
  & RL & Average & 8.83 & 21.67 & 22.67 & 1.0 \\
  & & Minimum & 7 & 27 & 27 & 1 \\
  \hline
\end{tabular}
\caption{Length and composition of constructed circuits. Results are based on transpiling top-performing circuits for the \texttt{IBM Heron} processor. GRL achieves smaller gate counts compared to RL-only.}
\label{tab:gate_counts}
\end{subtable}

\vspace{0.4cm}

\begin{subtable}[t]{0.8\textwidth}
\centering
\begin{tabular}{l|cccc|cccc}
\hline
\multirow{2}{*}{Backend name} 
& \multicolumn{4}{c|}{\textbf{3-qubit}} 
& \multicolumn{4}{c}{\textbf{2-qubit}} \\
\cline{2-9}
& \multicolumn{2}{c}{GRL} & \multicolumn{2}{c|}{RL} 
& \multicolumn{2}{c}{GRL} & \multicolumn{2}{c}{RL} \\
\cline{2-9}
& Avg. & Min. & Avg. & Min. & Avg. & Min. & Avg. & Min. \\
\hline
\texttt{fake\_torino}   & $\mathbf{-3.309}$ & $\mathbf{-3.366}$ & $-3.287$ & $-3.351$ & $\mathbf{-2.188}$ & $\mathbf{-2.213}$ & $-2.164$ & $-2.1992$ \\
\texttt{fake\_kawasaki} & $\mathbf{-3.370}$ & $\mathbf{-3.397}$ & $-3.235$ & $-3.319$  & $\mathbf{-2.162}$ & $\mathbf{-2.196}$ & $-2.123$ & $-2.1592$ \\
\texttt{fake\_quebec}   & $\mathbf{-3.318}$ & $\mathbf{-3.379}$ & $-3.266$ & $-3.318$  & $\mathbf{-2.118}$ & $\mathbf{-2.145}$ & $-2.084$ & $-2.1597$ \\
\hline
\end{tabular}
\caption{Performance comparison of GRL and RL agents for 3-qubit and 2-qubit TFIM ground state preparation at $h=1$.}
\label{tab:real-device-sim}
\end{subtable}

\caption{{GRL vs.\ RL on \texttt{IBM Heron} backends.} 
(a) Length and composition of constructed circuits after transpiling to \texttt{IBM Heron}.
(b) Average (Avg.) and minimum (Min.) energies over noisy-simulator evaluations on several fake backends.}
\label{tab:combined}
\end{table}

GRL's extended action space, which includes the \texttt{IBM Heron} processor's gateset and an additional gadget, likely contributes to its ability to find more optimal circuit configurations. Hence, to check this in supplementary fig.~\ref{fig:transpiled_univ}, we illustrate one of the best-performing circuits by the RL. Whereas, in supplementary fig.~\ref{fig:transpiled_grl}, we show the best circuit obtained for solving the same problem using GRL.

\begin{figure}[h!]
    \vskip 0.2in
    \begin{center}
        \centerline{\includegraphics[width=0.7\columnwidth]{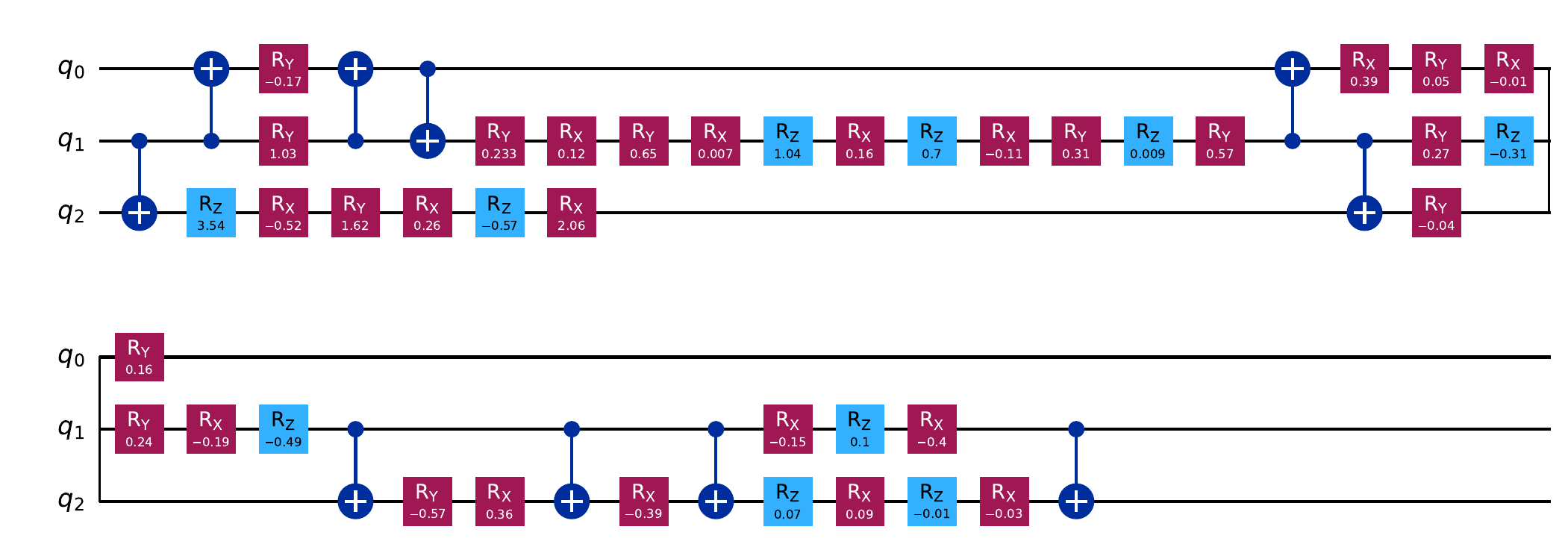}}
        \caption{{Best-performing circuit obtained from curriculum reinforcement learning (RL) agent in solving $N=3$ TFIM using a universal gate set}. We train the agent for $5000$ episodes and choose the circuit that provides the lowest error in estimating the ground state energy.}
        \label{fig:transpiled_univ}
    \end{center}
    \vskip -0.2in
\end{figure}
\begin{figure*}[h!]
    \centering
    \subfloat[\centering ]{{\includegraphics[width=0.6\textwidth]{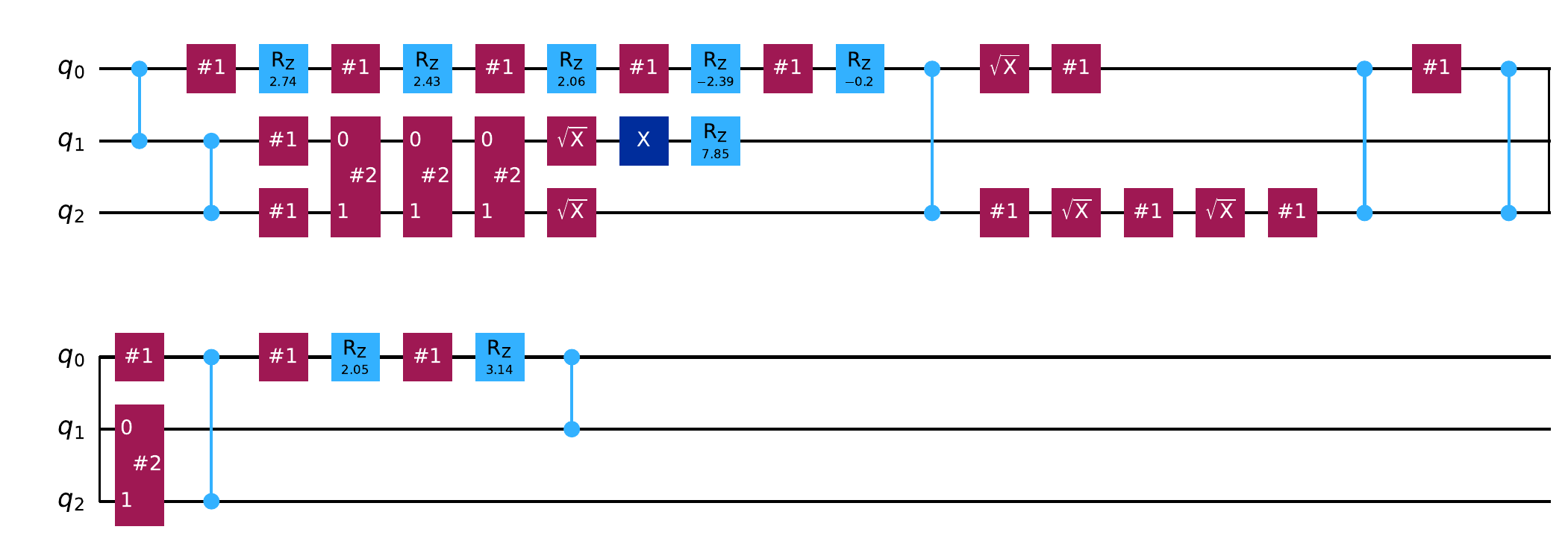} }}%
    \qquad
    \subfloat[\centering ]{{
                \includegraphics[width=0.2\textwidth]{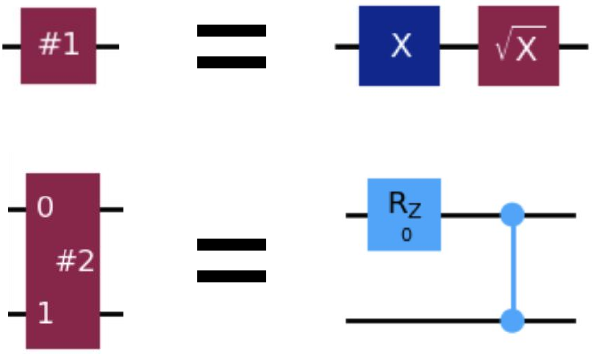} }

    }%
    \caption{(1) {Best-performing circuit obtained from the gadget reinforcement learning (GRL) agent with two gadgets in finding the ground state of a 3-qubit TFIM}. Similar to supplementary fig.~\ref{fig:transpiled_univ} we train the GRL agent for $5000$ episode and then choose the circuit that gives the lowest error in ground state estimation. In (2) we illustrate the extracted gadgets from easier problems with 2-qubit TFIM. }%
    \label{fig:transpiled_grl}%
\end{figure*}

Before implementation on real hardware, we would need to transpile the circuits to only use the instructions available on the specific platform.
Supplementary fig.~\ref{fig:transpiled_comparison} compares the transpiled circuit obtained through the RL agent with a universal gate set with that of our GRL agent with two extracted gadgets. We show a single example as an illustration.
\begin{figure*}[h!]
    \centering
    \subfloat[\centering ]{{\includegraphics[width=0.6\textwidth]{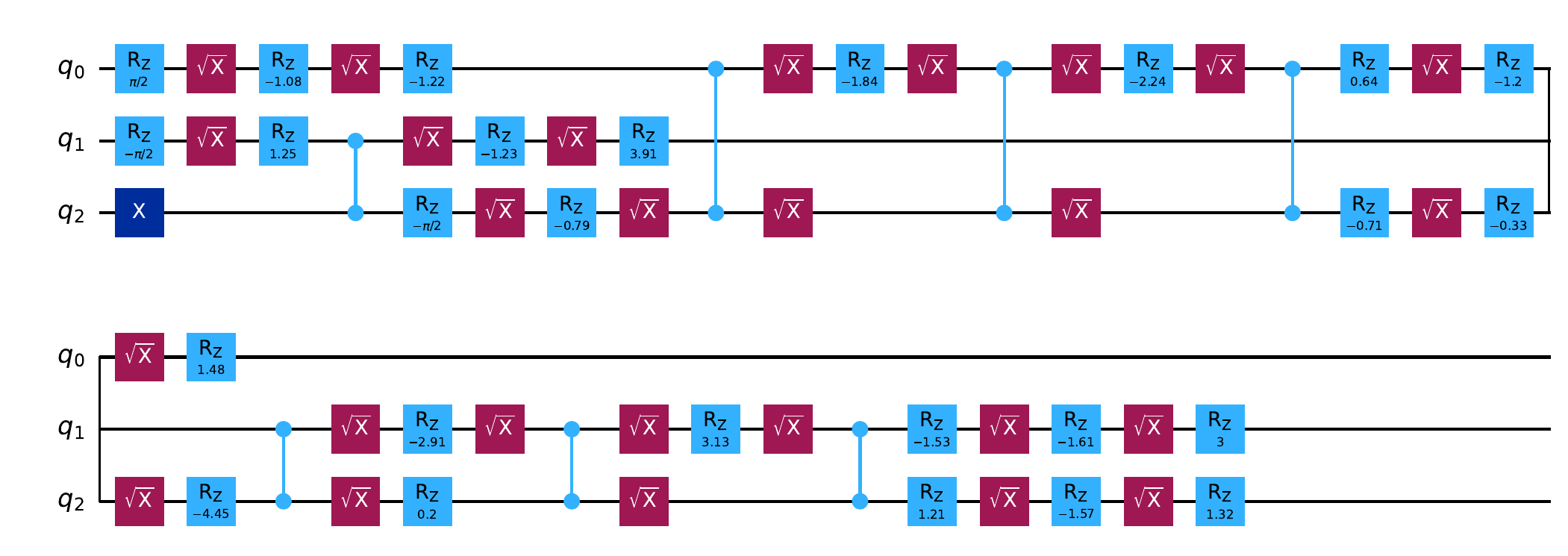} }}%
    \qquad
    \subfloat[\centering ]{{
                \includegraphics[width=0.7\textwidth]{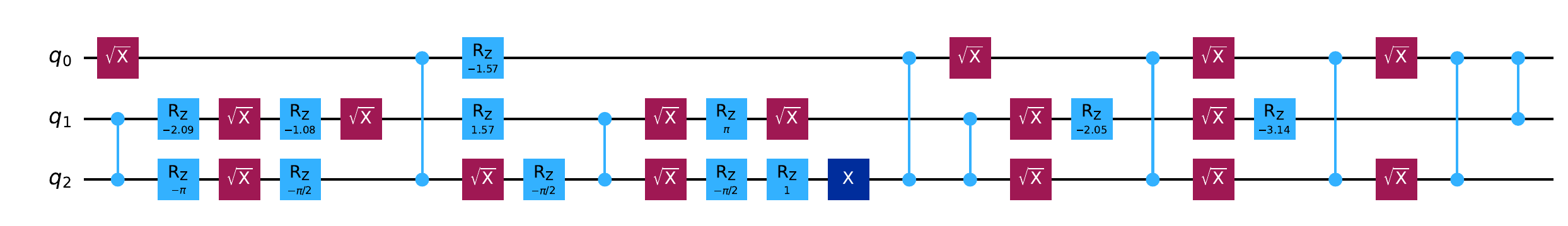} }

    }%
    \caption{{Comparison between the transpiled circuit obtained from (1) reinforcement learning (RL) using a universal gate set, (2) gadget reinforcement learning (GRL) using the native gateset for the \texttt{IBM Heron} processor and two gadgets}. After transpilation in real hardware, the circuit produced by GRL is more compact compared to the RL-agent circuit.}%
    \label{fig:transpiled_comparison}%
\end{figure*}

\subsection*{Supplementary Note 8.4: Scaling of GRL}
In supplementary fig.~\ref{fig:scalability-10qubit} we show that GRL with just one gadget can be extended to systems up to 10-qubit.
\begin{figure}[h!]
\centering\includegraphics[width=0.5\columnwidth]{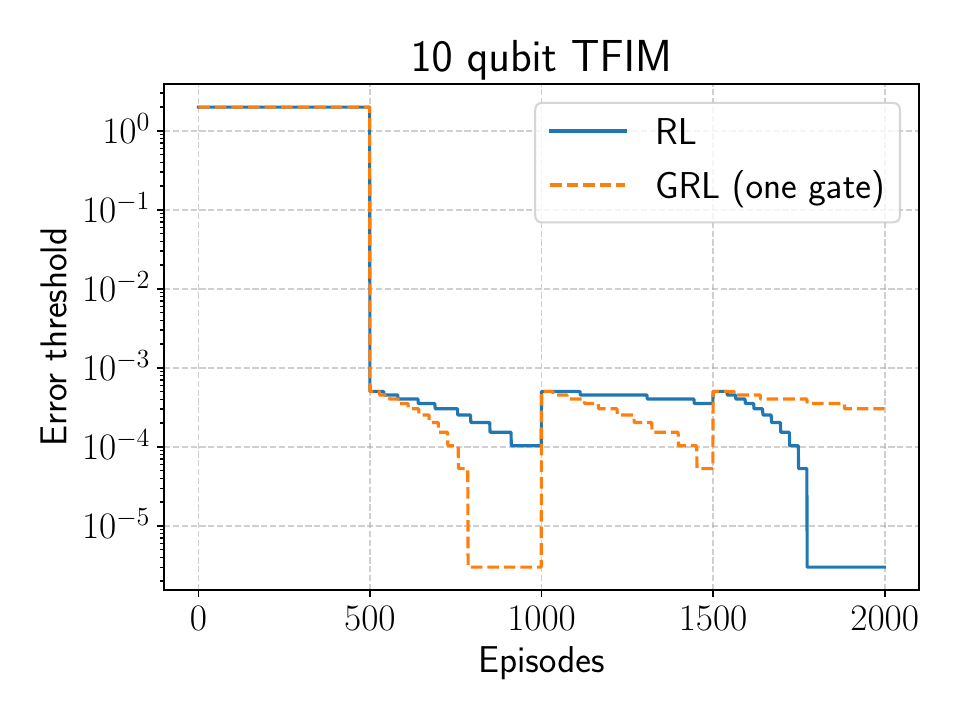}
    \caption{{Error threshold achieved for the 10-qubit TFIM using RL and GRL (with one gadget) as a function of episode number. Results demonstrate effective scalability, extending GRL capabilities to tens of qubits.}}
    \label{fig:scalability-10qubit}
\end{figure}

\subsection*{Supplementary Note 8.5: GRL for quantum chemistry}
In supplementary fig.~\ref{fig:supp_chem_bench}, we extend the application of GRL to a quantum chemistry problem where our main aim is to find the ground state of 3-qubit H$_2$ molecule. Furthermore we see that the gadgets learned by solving the H$_2$ molecule i.e. \texttt{CZ$_{ij}$RZ$_i$CZ$_{ij}$} and \texttt{X$_i$CZ$_{ij}$}) can be further utilized to accelerate the convergence towards even lower error.
\begin{figure}[h!]
    \centering
    \includegraphics[width=0.5\linewidth]{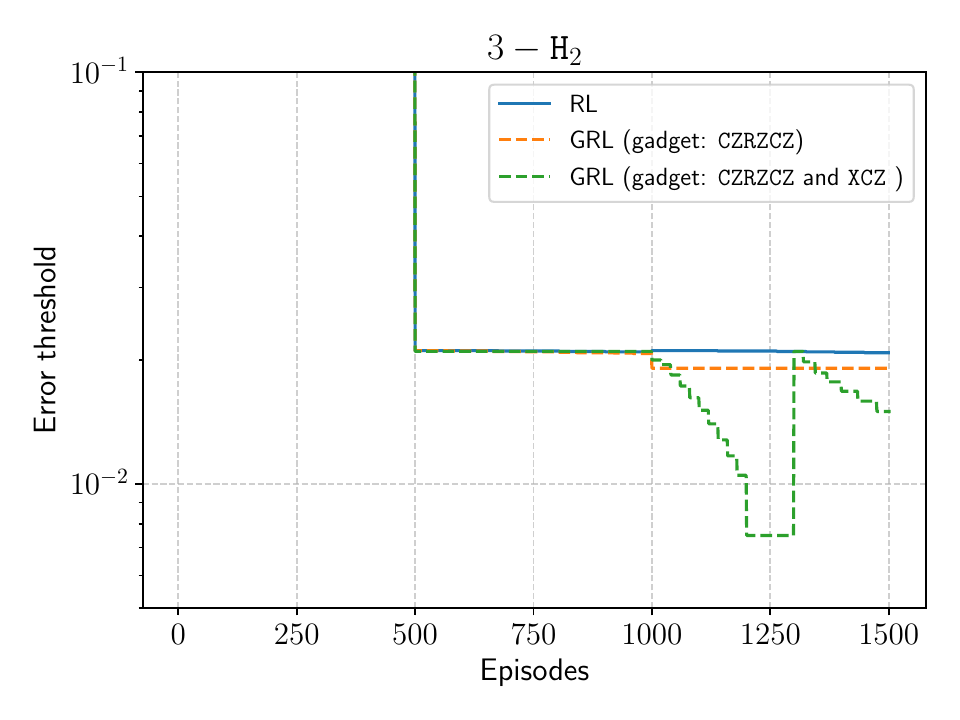}
    \caption{\small{{Transfer of gadgets to 3-qubit H$_2$.}
    We transfer gadgets (\texttt{CZ$_{ij}$RZ$_i$CZ$_{ij}$} and \texttt{X$_i$CZ$_{ij}$}) learned from smaller 2-qubit H$_2$ instances. GRL with transferred gadgets outperforms the standard RL baseline by achieving lower asymptotic error thresholds for 3-qubit H$_2$, with the combination of both gadgets yielding the largest improvement.}}
    \label{fig:supp_chem_bench}
\end{figure}

{\section*{Supplementary Note 9: The scalability of program synthesis }\label{appndix:scaling_program_synthesis}
Our proposed program synthesis approach demonstrates efficient scalability up to a predefined depth. We analyzed TFIM problems ranging from 2 to 5 qubits, as shown in supplementary tab.~\ref{tab:program_synthesis_performance_comparison}. The synthesis time increases with circuit depth and qubit count, reaching approximately 1.3 hours for 3-qubit TFIM circuits of depth 10.}

{While this scaling still suffers from the combinatoric explosion for deep quantum circuits, our results indicate that a circuit depth of up to 7 is sufficient for finding relatively simple gadgets and bootstrapping the search for subsequent iterations. Beyond this depth, the algorithm tends to discover more elaborate gadgets that encompass simpler ones. The key conclusions are as follows:
\begin{itemize}
\item {Time complexity}: Synthesis time ranges from 61.84 seconds for 2-qubit, depth-5 circuits to 4780.82 seconds for 3-qubit, depth-10 circuits.
\item {Gadget Extraction}: For our work, gadget extraction can take up to 2 hours. However, gadgets found using 2-qubit TFIM are sufficient to solve TFIM problems three times larger in terms of the number of qubits.
\item {Scalability}: The approach scales reasonably well for smaller qubit counts and depths but becomes more time-consuming for larger systems.
\item {Gadget complexity}: Simpler gadgets are found at lower depths, while more complex gadgets emerge at higher depths.
\end{itemize}}

{Our current implementation utilizes two gadgets throughout the work. Further research could explore the use of additional gadgets to potentially improve accuracy, for example for quantum circuit transpilation.}
\begin{figure}[t!]
    \centering
    \small
	\includegraphics[width=\columnwidth]{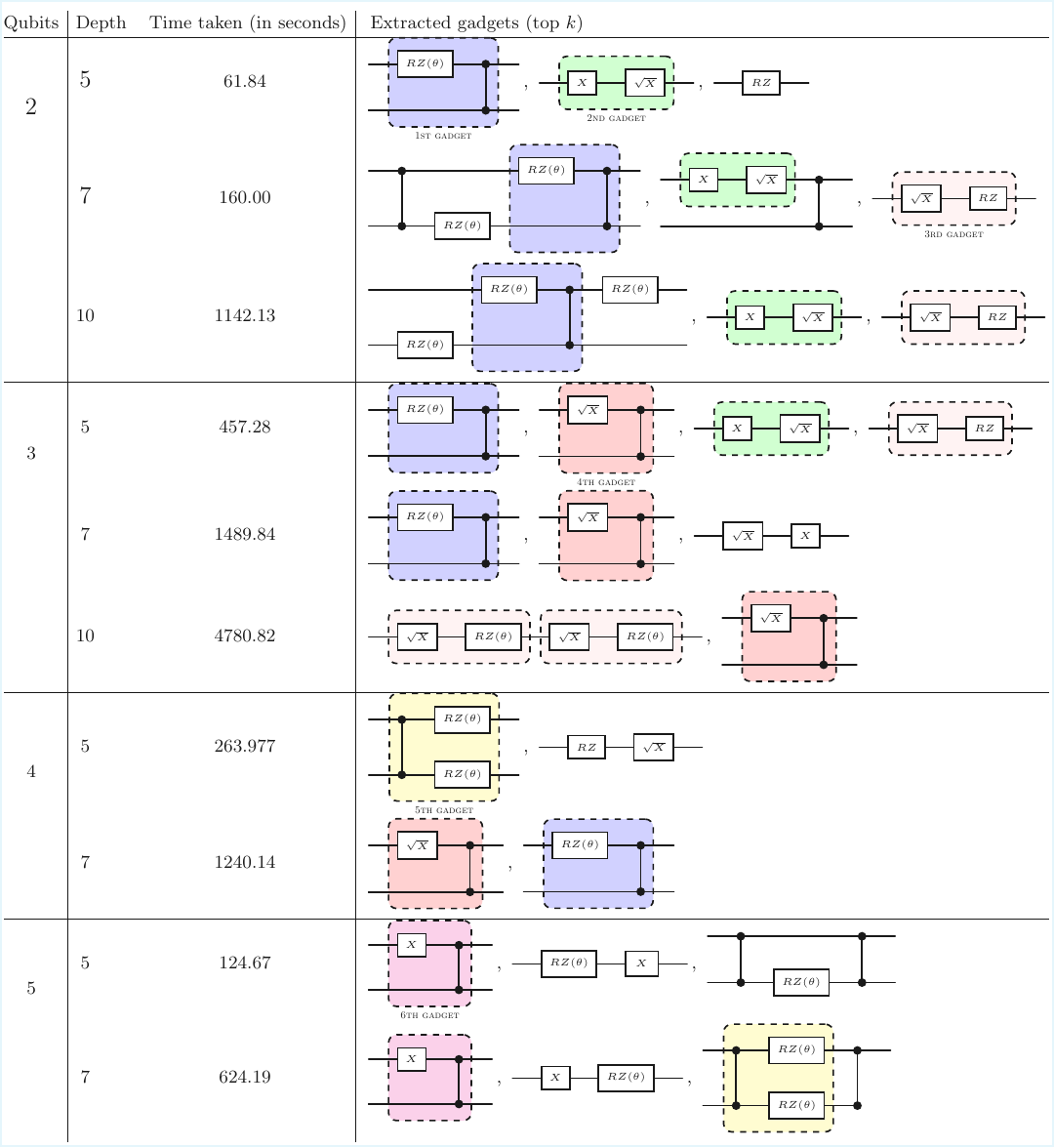}
    \caption{{\small {The scalability of program synthesis for $k$ circuits upto 5 qubits}. The circuits are obtained by solving $2$, $3$, $4$ and $5$-qubit TFIM with $h=10^{-3}$ and then sampling the top $10$ circuits sorted as per their accuracy towards ground state estimation.}}
    \label{tab:program_synthesis_performance_comparison}
\end{figure}

\section*{Supplementary Note 10: Analysis of training time}\label{app:training_time}
\begin{table}[t!]
\centering
\small
\begin{tabular}{c|cc|cc}
\hline
{Qubits} & \multicolumn{2}{c|}{{Training time (in hour)}} & \multicolumn{2}{c}{{Time to min error (in hour)}} \\
& RL & GRL & RL & GRL \\
\hline
2 & 6.3 & 4.5 & 5.8 & \cellcolor{green!30}2.2 \\
3 & 27.6 & 29.2 & 18.0 & \cellcolor{green!30}4.1 \\
4 & 29.5 & 29.8 & 11.1 & \cellcolor{green!30}5.0 \\
\hline
\end{tabular}
\caption{{{Gadget reinforcement learning (GRL) outperforms RL-only agents in finding the optimal solution more quickly for both the \( N=2 \), \( N=3 \) and \( N=4 \) qubit transverse field Ising model (TFIM).} The agent was trained with a fixed computational budget, equivalent to $5000$ episodes. The GRL agent identifies the optimal solution, much faster than the RL-only agent.}}
\label{tab:training_time}
\end{table}
In supplementary tab.~\ref{tab:training_time}, we compare the training time of reinforcement learning (RL) and gadget reinforcement learning (GRL). With a fixed computational budget of $5000$ episodes, the GRL agent identified the optimal solution, represented by a parameterized quantum circuit that approximates the TFIM ground state, much faster than the RL-only agent. 
This demonstrates GRL’s ability to achieve the desired accuracy with fewer interactions between the agent and the environment. 
This advantage makes GRL particularly effective in noisy environments. Moreover, by completing the task in less time, GRL significantly reduces energy consumption and computational resource requirements, making it a practical and efficient solution for resource-constrained scenarios.

{\section*{Supplementary Note 11: Impact of noisy gadgets}\label{appndx:impact_of_noise}
To simulate realistic quantum hardware conditions, we introduce a probabilistic Pauli noise channel after each controlled rotation operation. Specifically, after applying the first and the second gadget in GRL, we sample a random single-qubit unitary with probability $p = 0.5$ and apply it to the control qubit. The unitary is drawn from the Haar-random ensemble using
\begin{figure}[h!]
    \centering
    \includegraphics[width=0.5\linewidth]{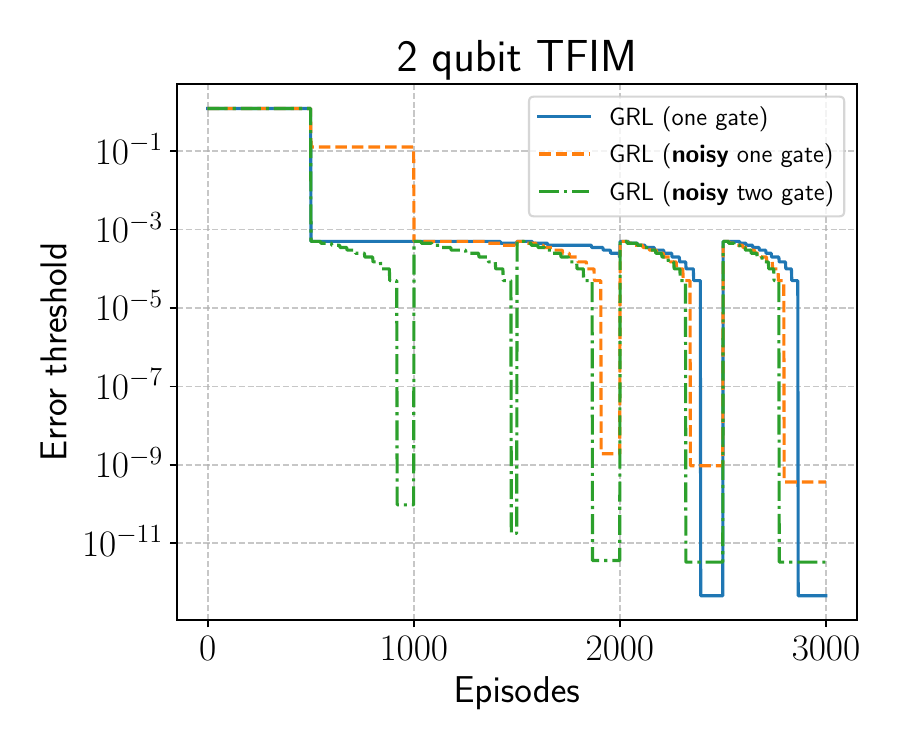}
    \caption{{{Comparison of GRL between a noiseless and noisy gate setting. } The impact of noise in GRL can be tackled by increasing the number of gadgets. As expected, the overall performance in providing a good approximation to the TFIM ground state decreases. Specifically, by using two gadgets instead of one, we can achieve similar performance with noisy gates to that obtained with one gadget with ideal gates.}}
    \label{fig:noise_effect}
\end{figure}
$\texttt{random\_unitary(2)}$ (from Qiskit~\cite{aleksandrowicz2019qiskit}), with a fixed seed for reproducibility. There are also other open-source frameworks, such as \texttt{QuantumCircuitOpt}~\cite{nagarajan2021quantumcircuitopt}, which can be utilized in modeling the noise. This noise model captures incoherent errors that may arise from imperfect control operations or environmental decoherence. By randomizing the noise unitary, we account for a broad class of single-qubit errors while maintaining tractability for numerical simulations.
\begin{itemize}
    \item {Probabilistic noise injection:} Noise is applied with $50\%$ probability per $\texttt{rzcz}$ operation.
    \item {Unitary noise:} General single-qubit errors are introduced via Haar-random unitaries.
    \item {Reproducibility:} A fixed seed ensures consistent benchmarking.
    \item {Physical motivation:} The model mimics control errors in near-term quantum devices.
\end{itemize}}

{Our analysis in supplementary fig.~\ref{fig:noise_effect} of the 2-qubit TFIM under noisy gate operations reveals a counterintuitive phenomenon: 
while one- and two-gate noise ({GRL (noisy one gate)} and {GRL (noisy two gate)}) initially degrade performance compared to the noise-free baseline, 
with more gadgets, the overall noise on the performance mitigates (as shown in the green line in supplementary fig.~\ref{fig:noise_effect}). 
This effect can be observed as a slower decay in the error threshold when scaling from one to two noisy gates, suggesting that increased gadget redundancy can partially absorb incoherent errors. This behavior is aligned with fault-tolerant design principles. The trajectory of performance decline flattens as more gadgets are incorporated, implying diminishing marginal damage from additional noise sources. These findings advocate for gadget architectures as a practical alternative to full error correction in near-term quantum reinforcement learning.}

\section*{Supplementary Note 12: IBM Heron processor}\label{app:heron_topology}
Supplementary fig.~\ref{fig:ibm_heron_processor} shows the topology of the \texttt{IBMQ Torino} platform.

\begin{figure}[h!]
    \centering
    \includegraphics[width=0.5\linewidth]{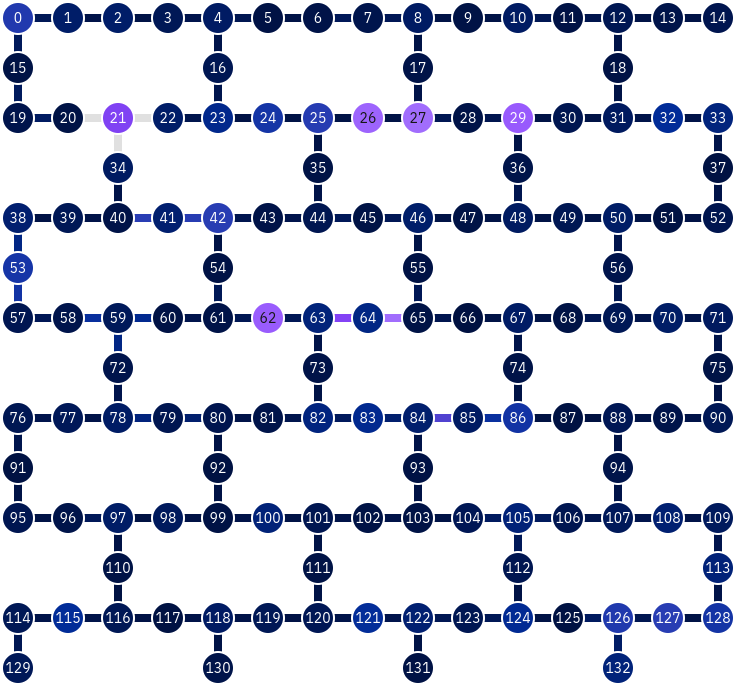}
    \caption{The topology of \texttt{IBMQ Torino}, which operates on \texttt{IBM Heron} processor.}
    \label{fig:ibm_heron_processor}
\end{figure}

\end{document}